\documentclass[10pt]{article} 
\usepackage{graphicx}
\usepackage{multicol}        
\usepackage{hyperref}
\usepackage{amssymb}
\usepackage{amsmath}
\usepackage{array}
\usepackage{graphics}
\usepackage{graphicx}
\usepackage{array}
\usepackage{fancyhdr}
\usepackage{enumerate}
\usepackage{floatrow}

\usepackage{authblk}
\title{An introduction to astrophysical observables in gravitational wave detections}
\author[1,2]{Maurizio Spurio}
\affil[1]{\scriptsize{INFN - Sezione di Bologna, Viale Berti-Pichat 6/2, 40127 Bologna, Italy}}
\affil[2]{\scriptsize{Dipartimento di Fisica e Astronomia dell'Universit\`a, Viale Berti Pichat 6/2, 40127 Bologna, Italy}}
\begin{document}
\maketitle

\begin{abstract}
Our knowledge and understanding of the Universe is mainly based on observations of the electromagnetic radiation in a wide range of wavelengths. Only during the past two decades, new kinds of detectors have been developed, exploiting other forms of cosmic probes: individual photons with energy above the GeV, charged particles and antiparticles, neutrinos and, finally, gravitational waves. These new ``telescopes'' leaded to unexpected breakthroughs. 
Years 2016 and 2017 have seen the dawn of the astrophysics and cosmology with gravitational waves, awarded with the 2017 Nobel Prize. The events GW150914 \cite{1} (the first black hole-black hole merger) and GW170817  \cite{2} (the coalescence of two neutron stars, producing a short gamma-ray burst and follow-up observed by more than 70 observatories on all continents and in space) represent really milestones in science that every physicist (senior or in formation) should appreciate. 

In this document, after an accessible discussion on the generation and propagation of GWs, the key features of observable quantities (the strain, the GW frequency $\nu_{gw}$, and $\dot \nu_{gw}$) of GW150914 and GW170817 are discussed using Newtonian physics, dimensional analysis and analogies
with electromagnetic waves. 
The objective is to show how astrophysical quantities (the initial and final masses of merging objects, the energy loss, the distance, their spin) are derived from observables. 
The results from the fully general-relativistic analysis published in the two discovery papers are compared with the output of our simple treatment.
Then, some of the outcomes of GW observations are discussed in terms of multimessenger astrophysics.

\end{abstract}


\section{Introduction}
\label{sec:intro}
On February 11, 2016 the LIGO collaboration announced the discovery of gravitational radiation due to the merger of a binary black hole 1.2 billion lightyear far from the Earth \cite{1}. 
This discovery represents a major scientific breakthrough for astrophysics, cosmology and even particle physics.
The 2017 Physics Nobel prize was awarded to R. Weiss, B. Barish and K. Thorne (all three members of the LIGO/Virgo Collaboration) for decisive contributions to the LIGO detector and the observation of gravitational waves (GWs).

Probably even more important, on October 16, 2017 the LIGO/Virgo collaboration announced \cite{2}, together with a large number of other experiments \cite{3}, the first observation of GWs and electromagnetic radiation. These observations are connected to a collision of two neutron stars 130 million lightyears far away from Earth producing a short gamma ray burst. The electromagnetic observations in the following days revealed signatures of recently synthesized material, including gold and platinum, solving a decades-long mystery of where about half of all elements heavier than iron are produced. 

Gravitational waves are ripples in the curvature of space-time, generated by accelerated masses that propagate from their production regions at the speed of light. After a long scientific discussion (\S \ref{sec:story}), they were deduced based on the Einstein's general theory of relativity (\S \ref{sec:phgw}). Gravitational waves transport energy in a form of radiant energy similar to electromagnetic radiation (\S \ref{sec:energy}).
In contrast to the incoherent superposition of emission from the acceleration of individual electric charges, GWs result from coherent, bulk motions of matter. 
Because they transfer very small amounts of energy to matter, GWs are able to penetrate the very densely concentrated matter that produces them. For this reason, on one hand, observations of GWs provide additional information for the study of high-energy processes in the Universe; on the other hand, their detection has represented a phenomenal challenge from the experimental point of view (\S \ref{sec:experiments}). 

In this document, key features of the observed gravitational radiation in the first observed binary Black Hole (BH) merger (GW150914, \S \ref{sec:gw15}) and in the first binary Neutron Star (NS) merger (GW170817, \S \ref{sec:gw1708}) are provided in terms of introductory physics. 
Data extracted from plots reported in the discovery papers are interpreted using Newtonian gravity, dimensional analysis and analogies with electromagnetic waves to make estimates of the astrophysical parameters. Key parameters obtained in this way (masses of merging objects, distances, emitted energy) are compared with the parameters reported in the discovery papers \cite{1,2,3} where they were extracted by fitting data to templates generated by numerical relativity.
A similar efforts have been carried out in \cite{LV,mathur}.

With few months of GW data, a catalog of binary BH merging events was produced and it is growing using the observing runs started in April 2019. In the near future, observations of GWs would potentially provide insights into topic astrophysical problems, as the formation of black holes via supernovae, binary interactions of massive stars, stellar cluster dynamics, and the formation history of black holes across cosmic time (\S \ref{sec:astrobh}). 

Finally, the experimental opportunity to relate observations of GW with traditional astrophysics observations, and the findings from charged cosmic rays, $\gamma$-rays, neutrinos is usually referred as \textbf{multimessenger astrophysics}. 
As demonstrated by GW170817 (see \S \ref{sec:GWcosmo}), observations made with different probes started to produce scientific breakthroughs of paramount importance.

Multimessenger astrophysics interconnects researchers with multivariate cultural background, spacing from particle physics, astrophysics, optics, general relativity, cosmology... 
My experimental activity is mainly related to neutrino physics and astrophysics. 
Both for research needs and for didactic activities, I tried to summarize the connections between particle physics and astrophysics in a book that (in the first edition) encompassed charged cosmic rays, $\gamma$-rays and neutrinos. 
The deep impact of GW150914 and GW170817 on this research field imposed a second edition \cite{spurio} that includes (in addition to a revision of the experimental results on charged particles and antiparticles; GeV and TeV $\gamma$-ray; neutrinos; dark matter) a new chapter devoted to GW observations and their integration with the other probes. 
This document contains (in a self-consistent way) most of the discussion included in Chapter 13 of \cite{spurio}. 
I hope this material can be useful to non-expert, young researchers and graduate students.

\section{Long story short}
\label{sec:story}

\subsection{``Are there any gravitational waves?''}

\textit{Gravitational waves} were firstly proposed in 1907 by the French physicist Henri Poincar\'e ({}``ondes gravifiques'') as emanating from massive bodies and propagating at the speed of light. 
The mathematical framework necessary for their description is that of the theory of general relativity, published afterwards in 1915.
Einstein himself, based on various approximations, derived three types of propagating solutions from the field equations, designed as longitudinal-longitudinal, transverse-longitudinal, and transverse-transverse oscillations.
However, the nature of Einstein's approximations led many (including Einstein himself) to doubt the result.
In 1922, Arthur Eddington showed that two types of wave were artifacts resulting from the choice of coordinate system (a sort of ``gauge effect''), and could be made to propagate at any speed by choosing appropriate coordinates. The famous Eddington's joking sentence that GWs {}``propagate at the speed of thought'' appears today in the title of a recommended monography \cite{thou} on the subject.
For the historical path toward a theoretical understanding of GWs, see also the recent \cite{cerva,cheng}. 

In 1936, Einstein and Nathan Rosen submitted a paper to the Physical Review Letter with the title \textit{``Are there any gravitational waves?''} 
The original version of the manuscript does not exist today, but Einstein's epistolary documents show that the answer to the title was {}``they do not exist''.
The editor sent the manuscript to be reviewed by an anonymous referee (in the usual peer review process), who questioned the conclusion of the paper (today, we know that the anonymous referee was Howard P. Robertson).
Einstein angrily withdrew the manuscript, asserting that he would never publish in the Physical Review again \footnote{The GW discovery paper \cite{1} was published on PRL!}. 
By some fortuitous circumstance, Leopold Infeld (at that time, an assistant of Einstein) met Robertson at a conference, the latter subsequently convincing Infeld that the conclusion in his presentation (that contained in the Einstein-Rosen paper) was incorrect.
Ultimately, Infeld similarly convinced Einstein that the criticism was correct; the paper was rewritten with the same title, the opposite conclusion and published elsewhere.

The question whether the waves carry energy (and are thus {}``physical'' objects) or are instead a {}``gauge'' effect remained controversial up to the end of the 1950s.
Finally, F. Pirani \cite{pirani} showed that gravitational waves would exert tidal forces on intervening matter, producing a pull and stretch in the material with a quadrupole oscillation pattern.
Contrarily to electromagnetic waves, man-made GWs cannot be produced. 
The only possibility to discovery them relies on the existence of very dense and massive astrophysical objects, as \textit{black holes} and \textit{neutron stars} (see \S \ref{sec:nsbh}).

These ideas stimulated experimental searches for gravitational radiation, which started in the 1960s with the work of Weber \cite{weber}. He began to speculate as to the way in which GWs might be detected, also motivated by incorrect predictions concerning the possibility of waves with amplitude (or \textit{strain}, a dimensionless quantity defined in the following sections) on the order $10^{-17}$ at frequencies near 1 kHz. 
At the University of Maryland, Weber built an aluminium bar 2 m in length and 0.5 m in diameter, with resonant mode of oscillation of $\sim 1.6$ kHz. The bar was fitted with piezo-electric transducers to convert its motion into an electrical signal. 
In 1971, with the coincident use of two similar detectors (the second was in Illinois), Weber claimed detection of GWs from the direction of the galactic center. This led to the construction of many other bar detectors of comparable or better sensitivity, which never confirmed his claims.

Improved theoretical models and calculations that appeared in the 1970s showed that gravitational wave strains were likely to be of the order on $10^{-21}$ or less and could encompass a wide range of frequencies. 
The correctness of such theoretical results remained a matter of controversy into the 1980s. The question would ultimately be solved by the observation of the Hulse-Taylor binary pulsar system: the rate of decrease of orbital period is 76.5 microseconds per year, in accord with the predicted energy loss due to gravitational radiation (\S \ref{sec:gw17}).
Thus, with respect to resonant bars, a more sensitive and wider-band detection technique was necessary. Such a technique became available with the development of laser interferometers. 
After the prototype demonstrations at Caltech, Glasgow, and Garching, funding agencies in the USA and Europe committed to the construction of large, kilometer-scale laser interferometers: LIGO (USA, 4 km), Virgo (France and Italy, 3 km) and GEO (UK and Germany, 600 m). The length of their arms today allows for a strain sensitivity on the order of $10^{-22}$ over a 100 Hz bandwidth, a development that finally led to the discovery in 2015.

\subsection{The main characters: Neutron Stars and Black Holes}
\label{sec:nsbh}

A \textbf{neutron star (NS)} is the result from the supernova explosion of a massive star, combined with gravitational collapse, that compresses the core to the density of nuclear matter ($\sim 4\times 10^{17}$ kg/m$^3$).
Neutron stars are supported against further collapse by neutron degeneracy pressure, a phenomenon described by the Pauli Exclusion Principle. If the remnant star has a mass greater than $\sim 3$ solar masses (the solar mass is shortened in following with the symbol $M_\odot\simeq 1.99\times 10^{30}$ kg), it continues collapsing to form a black hole. 
The maximum observed mass of neutron stars is $\sim 2.1 M_\odot$.
Typically NSs can have masses of $\sim 1.4 M_\odot$ and, at the nuclear matter density, they have radius of the order of 10 km. 

The estimated number of NSs in our Galaxy is ${\mathcal O}(10^8)$, a figure obtained from the number of stars that have undergone supernova explosions. 
Most of them are old and cold, and NSs can only be easily detected if they are young, rotating systems (in this case, they are usually referred for as \textit{pulsars} \footnote{About 2000 pulsars are present in the catalogue available at \url{http://www.atnf.csiro.au/people/pulsar/psrcat/}. }
or part of a binary system.
Presently eight binary NS systems are known in our Galaxy, including the Hulse-Taylor binary. 

As a first approximation, NSs are composed entirely of neutrons; the electrons and protons present in normal matter have combined in the collapsing phase to produce neutrons. 
However, current models indicate a possible onion-like structure.
The surface of a NS should be composed of ordinary atomic nuclei crushed into a solid lattice, together with a plasma of electrons. 
Due to their high binding energy per nucleon, iron nuclei could be predominant at the surface, or immediately under the surface made of lighter nuclei.
Proceeding inward, nuclei with ever-increasing numbers of neutrons would be present; such nuclei, if free, would decay quickly, but they are kept stable by enormous pressures. Then, the concentration of free neutrons increases rapidly until the core. 
The equation of state for a neutron star is still not known, in particular we do not know if the presence in the core of exotic forms of matter is allowed. These forms include degenerate strange matter (containing strange quarks in addition to up and down quarks), matter containing high-energy pions and kaons in addition to neutrons, or ultra-dense quark-degenerate matter.
As we discuss later, observations of binary NS merger would provide insight on their equation of state.

\vskip 0.2cm

\noindent A \textbf{black hole (BH)} is a massive object exhibiting such strong gravitational effects that nothing (particles and electromagnetic radiation) can escape from inside its boundary, called the \textit{event horizon}. In most cases, we can consider the event horizon equivalent to the \textit{Schwarzschild radius}. This is correct for non-rotating massive objects that fit inside this radius.

The escape velocity, $v_{esc}$ from a body of mass $M$ at a distance $r$ from the center (assuming that $r\ge R_b$, with $R_b$ the radius of the spherical body) is $v_{esc}=\sqrt{2GM/r}$.
The {Schwarzschild radius}, ${\mathcal R}$, is defined as the dimension of an object of mass $M$ such that $v_{esc}=c$. Using the above relation, we obtain:
\begin{equation}\label{eq:rd1}
{\mathcal R} = \frac{2GM}{c^2} = 2.95 \biggl(\frac{M}{M_\odot}\biggr) \ \textrm{ km}  \ ,
\end{equation}
quantity that scales linearly with the object mass.
If the body is sufficiently dense and confined within $\mathcal R$, the Schwarzschild radius represents its {event horizon} and its inner region behaves as a BH. Particles and light can escape the BH only if they remain outside the event horizon.

Although Eq. (\ref{eq:rd1}) is obtained from Newtonian considerations, the same conclusion emerges from general relativity. Furthermore, in classical general relativity, a particle that is inside the event horizon can never emerge outside.
More generally, BHs are particular solutions to the Einstein field equations (\S \ref{sec:phgw}). It has been demonstrated (by the so-called \textit{no-hair theorem}), that stable BHs are completely described at any time by the following quantities: 
$\textit{i)}$ the mass-energy, $M$;
$\textit{ii)}$ the angular momentum, or spin, $\vec{S}$ (three components);
$\textit{iii)}$ the total electric charge, $Q$.
In terms of these properties, four types of black holes can be defined: Uncharged non-rotating BHs (also called \textit{Schwarzschild BHs}) and rotating BHs (called \textit{Kerr BHs}). Then, there should be also charged non-rotating and rotating BHs. 
A rotating BH is formed in the gravitational collapse of a massive spinning star or from the collapse of a collection of stars or gas with a total non-zero angular momentum.
A rotating BH can loss rotational energy through different mechanisms occurring just outside its event horizon. 
In that case, it gradually reduces to a Schwarzschild BH, the minimum configuration from which no further energy can be extracted.

At present, the number of existing BHs in our Galaxy, their number density in the Universe, the mass spectrum, the existence of a gap in the mass spectrum in the range of 5-10 $M_\odot$, the upper limit of stellar BH masses, etc., are waiting for experimental inputs. They can be probably provided only through observations of the produced GWs in merging events.

\section{From Einstein Equation to Gravitational Waves}
\label{sec:phgw}

Space-time is considered (in general relativity) as a four-dimensional manifold, and gravity is a manifestation of the manifold's curvature. We recall here some fundamental concepts from general relativity, remanding a detailed description to more specialized texts (see, for instance, \cite{maggiore,saulson,hartle}).
The differential line element $ds$ at space–time point $\textbf{x}$ has the form:
$$ds^{2}=g_{\mu\nu}(\textbf{x})dx^{\mu}dx^{\nu}  $$
where $ g_{\mu\nu}$  is the symmetric metric tensor, and repeated indices imply summation. For a flat Cartesian coordinate metric $[\mu = (ct, x, y, z)]$:
\begin{equation}\label{eq:phgw02}
g^{Cart}_{\mu\nu}(\textbf{x})= 
{\begin{pmatrix}
-1&0&0&0\\
0&1&0&0\\
0&0&1&0\\
0&0&0&1\\
\end{pmatrix}}
\end{equation}
If the space is not \textit{flat}, the form of the metric tensor is much more complicated. 

Starting from the observed equivalence of gravitational and inertial mass, which was elevated to the status of a fundamental physical principle, Einstein interpreted gravity as the physical manifestation of curvature in the geometry of space-time.
The mathematical way adopted in the general relativity theory is to quantify the curvature of a metric via the covariant equation of motion for a test particle. Thus, space-time curvature is associated with matter and energy:
\begin{equation}\label{eq:phgw1}
G_{\mu \nu }\equiv R_{\mu \nu }-{\textstyle 1 \over 2}R\,g_{\mu \nu} ={8\pi G \over c^{4}}T_{\mu \nu } \ .
\end{equation}
On the left-hand side, $ G_{\mu \nu }$  is the Einstein tensor, which is formed from the Ricci curvature tensor  $ R_{\mu \nu }$ and the space-time metric $ g_{\mu \nu }$;  the matrix $G_{\mu \nu }$ is symmetric, and $R=g^{\mu \nu }R_{\mu \nu }$ is called the curvature scalar.
On the right-hand side, $ T_{\mu \nu}$ is the stress-energy tensor of matter fields, and $G$ is Newton's gravitational constant.  Equation (\ref{eq:phgw1}) derived by Einstein quantifies how energy density leads to curvature and, in turn, how curvature influences energy density.
Though simple in appearance, the Einstein equation is a nonlinear function of the metric and its first and second derivatives; this very compact geometrical statement disguises 10 coupled, nonlinear partial differential equations. 

In order to give a very simple mechanical analogy of (\ref{eq:phgw1}), consider the potential energy connected with the spatial deformation of a spring: 
\begin{equation}\label{eq:phgw2}
k x = \nabla U \ .
\end{equation}
Here, $x$ takes the place of the metric tensor and $U$ that of the stress-energy tensor. Thus, the equivalent of the spring's constant $k$ in (\ref{eq:phgw1}) is
\begin{equation}\label{eq:phgw3}
k \longrightarrow \quad \frac{c^4}{8\pi G}= 5.6\times 10^{45} \textrm{ kg m s}^{-2}
\end{equation}
This is equivalent to say that the energy required to distort space is analogous to that required to induce an elastic deformation of rigid materials, but to a much greater degree because space is extremely stiff. 

\begin{figure}[tb]
\begin{center}
\includegraphics[width=10.0cm]{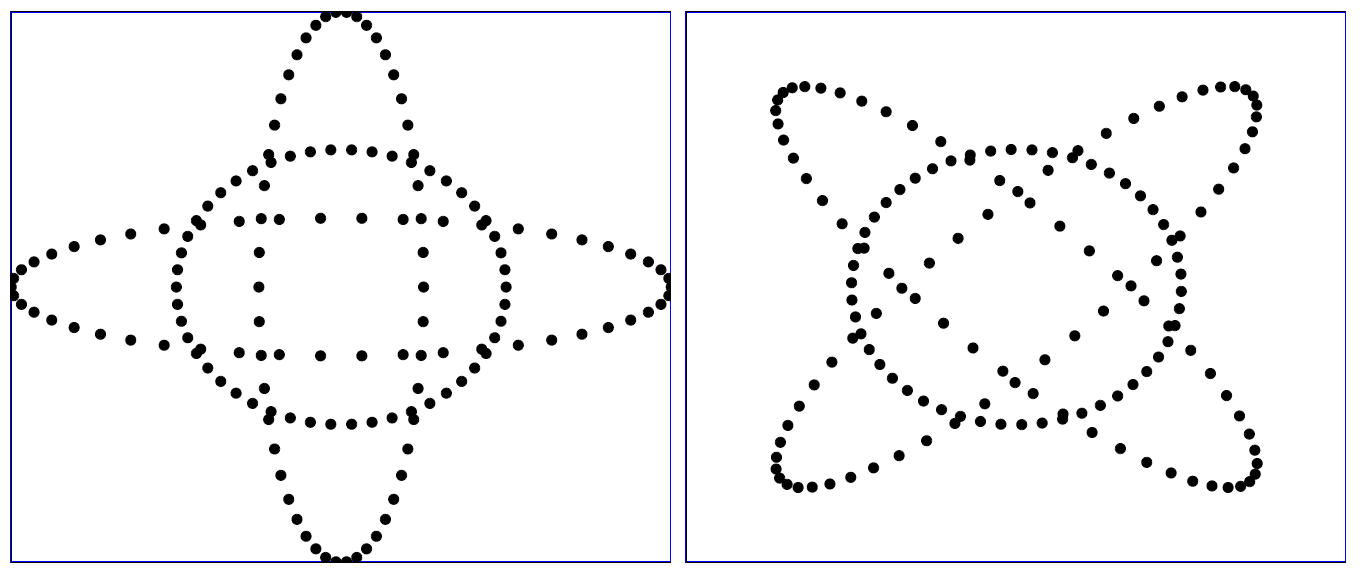}
\end{center}
\caption{\small\label{fig:h+hx} 
In the weak field, the gravitational waves have two independent polarizations called $h_+$ and $h_\times$. 
The effect on the separations of test masses displaced in a circular ring in the $(x, y)$-plane, perpendicular to the direction of the wave, is shown on the left for the $h_+$ wave and on the right for $h_\times$. The ring continuously gets deformed into one of the ellipses and back during the first half of a gravitational wave period and gets deformed into the other ellipse and back during the next half.}
\end{figure}

Generation of GWs is implicit in the Einstein equations. 
In fact, if we consider a small and flat region far from a non-static source (for instance, two massive objects orbiting each other), the gravitational field should vary with time. 
This can be thought as an effect of a GW that perturbs the flat Cartesian metric by only a small amount, $h_{\mu\nu}$:
\begin{equation}\label{eq:phgw4}
g_{\mu\nu}(\textbf{x})= g^{Cart}_{\mu\nu} + h_{\mu\nu}(\textbf{x})
\end{equation}
Under these assumptions, the left side of the Einstein equation (\ref{eq:phgw1}) can be greatly simplified by keeping only first order terms in $ h_{\mu\nu}$ and applying a gauge condition analogous to that applied on the electromagnetic potential. 
The choice of a particular gauge (gauge fixing) denotes the mathematical procedure for coping with redundant degrees of freedom in field variables
\footnote{
In the electromagnetic theory, the Lorenz gauge condition (or Lorenz gauge) is a partial gauge fixing of the four-vector potential. 
The condition is that $\partial _{\mu }A^{\mu }=0$. In ordinary vector notation and SI units, the gauge condition is written as 
$ \nabla \cdot {\vec {A}}+{\frac {1}{c^{2}}}{\frac {\partial\varphi}{\partial t}}=0$. 
This does not completely determine the gauge: one can still make a gauge transformation $A^{\mu }\to A^{\mu }+\partial ^{\mu }f$, where $f$  is a scalar function satisfying $ \partial _{\mu }\partial ^{\mu }f=0$.
}.
In vacuum ($T_{\mu\nu}$ = 0), one obtains the homogeneous wave equation:
\begin{equation}\label{eq:phgw5}
\biggl(-\frac{1}{c^2}\frac{\partial^2}{\partial t^2}+\nabla^2\biggr)h_{\mu\nu} (\textbf{x}) \equiv \Box h_{\mu\nu} (\textbf{x}) =0
\end{equation}
that has familiar space and time dependence solutions, for example for a fixed wave vector $\vec{k}$:
\begin{equation}\label{eq:phgw6}
 h_{\mu\nu}(\textbf{x})= h^0_{\mu\nu} e^{[i(\vec{k}\cdot\vec{x} -\omega t)]} \ ,
\end{equation}
but describes a tensor perturbation. The constant $ h^0_{\mu\nu}$ is a symmetric $4 \times 4$ matrix and $\omega = kc$. 
A particular useful solution for the GW in vacuum is obtained by choosing the $z$ axis along the direction of the wave vector $\vec{k}$; this condition is known as the \textit{transverse-traceless (TT) gauge} and leads to the relatively simple form:
\begin{equation}\label{eq:phgw7}
 h_{\mu\nu}(\textbf{x})= {\begin{pmatrix}
0&0&0&0\\
0&h_+&h_\times&0\\
0&h_\times&-h_+&0\\
0&0&0&0\\
\end{pmatrix}}
e^{[i(\vec{k}\cdot\vec{x} -\omega t)]} \ ,
\end{equation}
where $h_+$ and $h_\times$ are constant amplitudes.
For illustration, Fig. \ref{fig:h+hx} depicts the nature of these two polarizations as gravitational waves propagating along the $z$-axis impinge upon a ring of {}``free'' test masses in a plane perpendicular to the wave direction $\vec{k}$.

Eq. (\ref{eq:phgw7}) can be used to explain the effect of a GW impinging on free-fall test masses of a detector on Earth.
We need now to determine the relation of GWs to their source.
This is defined by the inhomogeneous Einstein equation (\ref{eq:phgw1}). Under the assumptions of a weak field in a nearly flat space-time, Cartesian coordinates and the transverse-traceless gauge, one has an inhomogeneous wave equation:
\begin{equation}\label{eq:phgw8}
\Box h_{\mu \nu }(\textbf{x}) ={16\pi G \over c^{4}}T_{\mu \nu } \ .
\end{equation}
This source equation is analogous to the wave equation originating from a relativistic electrodynamic field:
\begin{equation}\label{eq:phgw9}
\Box A^{\mu}(\textbf{x}) = - \mu_0 J^\mu  \ ,
\end{equation}
where $A^\mu = (\phi/c, {\vec A)}$ is the four-vector with the scalar and vector potential functions and $J^\mu = (c\rho ,{\vec J})$ that with the electric scalar charge and current density. 
In the case of electrodynamics, the Green function formalism is applied to derive the solution: for instance, the vector potential is written as an integral over a source volume: 
\begin{equation}\label{eq:phgw10}
{\vec A}^{\mu}(t,\vec{x}) = \frac{\mu_0}{4\pi} \int d^3x^\prime
\frac{[\vec{J}(t^\prime, \vec{x}^\prime)]_{ret}}{|\vec{x}-\vec{x}^\prime|}  \ ,
\end{equation}
where $[...]_{ret}$ indicates evaluation at the \textit{retarded time}  $t^\prime \equiv t - |\vec{x}-\vec{x}^\prime|/c$. 
Similarly, the solution for the waves (\ref{eq:phgw8}) produced by variations of the mass configuration can be written as
\begin{equation}\label{eq:phgw11}
h_{\mu\nu}(t,\vec{x}) = \frac{4G}{c^4} \int d^3x^\prime
\frac{[T_{\mu\nu}(t^\prime,\vec{x}^\prime)]_{ret}}{|\vec{x}-\vec{x}^\prime|}  \ ,
\end{equation}
In the following, we are interested to some particular solutions, namely that originated by a source with scale dimension $R$ that varies harmonically with time with characteristic frequency $\nu_{s}$, wavelength $\lambda=c/\nu_{s}$ and with the energy tensor dominated by the rest mass of the rotating objects.
This includes the systems with two massive objects (two black holes, or two neutron stars, or a black hole and a neutron star) orbiting around one another.
In addition, we assume that: 

\noindent \textbf{1)}  $\lambda \gg R$, i.e. the \textit{long-wavelength approximation}, and

\noindent \textbf{2)} $r\gg R$, where $r$ is the distance of the observer from the source (the \textit{distant-source approximation}). 

Under these approximations, the connection (\ref{eq:phgw11}) between the tensor $h$ and source reduces to
\begin{equation}\label{eq:phgw12}
h_{\mu\nu}(t,\vec{x}) \simeq \frac{4G}{rc^4} \int d^3x^\prime
\ T_{\mu\nu}(t- {r}/{c},\vec{x}^\prime) \ .
\end{equation}
This relation further simplify if we assume that the energy density of the source is dominated by its rest-mass density $\rho_m$ (non-relativistic internal velocities), obtaining a relation for the spatial coordinates:
\begin{equation}\label{eq:phgw13}
\boxed{
h_{ij} \simeq \frac{4G}{rc^4} \frac{d^2Q_{ij}}{dt^2}
\ .  }
\end{equation}
where $Q_{ij}$ is a $3\times 3$ tensor of the mass quadrupole moment: 
\begin{equation}\label{eq:phgw14}
Q_{ij}= \int d^3{{x}}  \biggl(x_i x_j -\frac{1}{3}r^2 \delta_{ij}  \biggr) \rho_m(\vec{x})
\end{equation}
Here, $\delta_{ij}$ is the Kronecker-delta matrix (diagonal elements =1, off-diagonal elements =0). Although $h_{ij}$ is a tensor quantity, in the following we indicate with $h$ the order-of-magnitude of its elements, i.e. the \textit{effect} of the GW. 

A tensorial object similar to (\ref{eq:phgw14}) appears in advanced courses of electromagnetism in the multipole expansions of charge distributions. It is simple to introduce it if you have familiarity with the moment of inertia tensor, $I$, introduced in mechanics (see for instance \cite{fey,ber}).
For a system of $n$ particles with masses $m_\alpha$ and positions $(x_\alpha, y_\alpha, z_\alpha)$, the elements of $I$ are:
\begin{equation}
I_{xx}= \sum_{\alpha=1}^n m_\alpha (y^2_\alpha +z^2_\alpha)\ ; \quad
I_{xy} = -\sum_{\alpha=1}^n m_\alpha x_\alpha y_\alpha \ , 
\end{equation}
and the other diagonal and off-diagonal components can be written down by analogy. The quadrupole tensor is similar: the off-diagonal components have the form $Q_{ij}=- I_{ij}$ and the diagonal components $Q_{xx}= -I_{xx}+(1(3) I$, and similarly for $Q_{yy}, Q_{zz}$. 
Here $I= I_{xx}+ I_{yy}+ I_{zz}$.

\section{Energy carried by a gravitational wave}
\label{sec:energy}

To summarize the content of the previous section, the effect of accelerated charges is to produce an electromagnetic wave with oscillating electric and magnetic fields propagating at the speed of light, $c$. 
The connection between sources ($J^\mu = (c\rho ,{\vec J})$, the electric scalar charge and current density) and potential ($A^\mu = (\phi/c, {\vec A)}$, the scalar and vector potential functions) is given by Eq. \ref{eq:phgw9}. The electric, $\mathbf E$, and magnetic, $\mathbf B$, fields are obtained from space-time derivatives of the potential. 
The effect of accelerated matter is to produce a GW that distort the local metric. In this case, the connection between sources (masses) and potential $h$\footnote{As common in the literature, we shorten for simplicity with $h$ both the $h_+$ and $h_\times$ degrees of freedom.} is provided by Eq. \ref{eq:phgw8}. The effect of a gravitational ripple is that the distance $L$ between two free masses can be stretched or shrunk by a quantity $\Delta L$ such that $h= \Delta L/L$. The quantity $\Delta L$ changes with time and the time derivative of the strain $h$, denotes as $\dot{h}$, is the gravitational equivalent of the electromagnetic field.  
The GWs propagate at the speed of light, as implicit in the $\Box$ operator defined in Eq. \ref{eq:phgw5}. 

To lowest order, gravitational radiation is a quadrupolar phenomenon.
In electromagnetism, radiation induced by electric dipole and magnetic dipole processes is supported, while ``monopole'' radiation is prohibited by electric charge conservation.
``Monopole'' gravitational radiation is prohibited by energy conservation; dipole radiation is related to the source's center of mass; momentum conservation ensures that a closed system's center of mass cannot accelerate and, correspondingly, there is no dipole contribution to GWs.
Note that, as for electrodynamics, gravitational radiation intensity is not spherically symmetric (isotropic) about the source.

The problem on how small is $h$, which are typical oscillation frequencies and which methods should be used to experimentally observe $\Delta L$ are the subjects of the following sections. 
Here, we concentrate on the problem of energy carried out by a GW.

As we mentioned before, a long discussion took place in the community about the energy flux implicit in GWs.
The computation is not easy, and we report only the salient results. 
The evaluation of the GW energy flux is easier if considered in a spatial volume encompassing many wavelengths, but small in dimension compared to the characteristic radius of curvature of the space. 
Under this assumption, the GW energy flux corresponds to: 
\begin{equation}\label{eq:enegw1} 
\boxed{ {\mathcal F}= \frac{1}{32\pi} |\dot{h}|^2  \frac{c^3}{G} \ .}
\end{equation}
The SI unit of the $\mathcal F$ vector is the Watt per square meter (W/m$^2$). 
It has the same units as the electromagnetic Poynting vector, $ \mathbf {S} ={\frac {1}{\mu _{0}}}\mathbf {E} \times \mathbf {B}$. 
The Poynting vector represents the directional energy flux (the energy transfer per unit area per unit time). 
We do not derive (\ref{eq:enegw1}) (see for instance \cite{maggiore}); however, it is easy to verify that the quantity $c^3/G$ has dimensions of [Energy Time/Area]; 
the quantity $|\dot{h}|$ [Time$^{-1}$] takes the place of the derivative of the electromagnetic potential, i.e. the electric and magnetic fields, and thus $(c^3 |\dot{h}|^2/G)$ with dimensions of [Energy /(Area Time)] has the role of $|\mathbf {S}|$. Finally, the numerical term $1/32\pi$ is the results on a heavy computation. 

As a general result \cite{maggiore,saulson,hartle} for the total luminosity (in Watt) of GWs in the radiation zone, ${\mathcal L}$ depends on the third time derivative of the mass quadrupole moment averaged over several cycles: 
\begin{equation}\label{eq:enegw3}
\boxed{
{\mathcal L}=\frac{1}{5} \frac{G}{c^5} 
\sum_{i,j=1}^3
\frac{d^3 Q_{ij}}{dt^3}
\frac{d^3 Q_{ij}}{dt^3} \ .
}
\end{equation}

In the following sections, we specify the above general formulas to the case of two-body systems. With some approximations, we can produce simple and reasonably accurate predictions for the frequency, duration, and strength of gravitational radiation from astrophysical sources.
Before turning to this, it is useful to consider some additional comparisons between the gravitational radiation and the electromagnetic radiation.

\vskip 0.1cm \noindent $\bullet$ In most astrophysical cases, emitted electromagnetic radiation is an incoherent superposition of light from sources much larger than the radiation wavelengths; in contrast, gravitational radiation likely to be detectable (below few kHz) comes from systems with sizes $R$ smaller (or, in some cases, comparable) to the emitted wavelength $\lambda$. Hence the signal reflects the {coherent motion} of extremely massive objects.

\vskip 0.1cm \noindent $\bullet$ Solutions of Maxwell's equations for a localized oscillating source of dimension $R$ at a distance $r$ in a homogeneous material (e.g., vacuum or air), are \textbf{E} and \textbf{B} fields that decay as $\frac{1}{r}$ when $r \gg R$. 
These conditions refer to the \textit{radiating fields}, and the condition $r\gg R$ defines the \textit{far field}. Similarly, the quantity describing the the strain, Eq. (\ref{eq:phgw13}), and its time derivative, $\dot h$, decrease as $\frac{1}{r}$.

\vskip 0.1cm \noindent $\bullet$ Detectors of the electromagnetic radiation are sensible to the flux intensity (i.e. to the Poynting vector, 
$ \mathbf {S} ={\frac {1}{\mu _{0}}}\mathbf {E} \times \mathbf {B}$), which decreases as $\frac{1}{r^2}$. 
This, because work must be done on electric charges (for example, in an antenna). 
The sensitivity of a detector represents the minimum magnitude of input signal required to produce a specified output signal.
Using an electromagnetic receiver with sensitivity $s_{EM}$, a given source of luminosity $L_o$ can be detected up to a maximum distance, ${r_{EM}}$, such that
${s_{EM}}\propto L_o/r^2_{EM}$. The number of detectable EM sources is proportional to the observable sky volume, $V_{EM}\propto r_{EM}^3$.

\vskip 0.1cm \noindent $\bullet$ GW interferometer detectors register waves coherently by following the \textit{phase} of the wave and not just measuring its intensity. 
The phase of the wave is contained in the strain $h$ that decreases as $\frac{1}{r}$. A GW receiver with sensitivity $s_{GW}$, a given source of luminosity $L_o$ can be detected up to a maximum distance, ${r_{GW}}$, such that
${s_{GW}}\propto L_o/r_{GW}$. The number of detectable GW sources is proportional to the observable sky volume, $V_{GW}\propto r_{GW}^3$.

\vskip 0.1cm \noindent $\bullet$ 
Let now consider a sensitivity improvement of a factor of $k$ in the EM detector, i.e.,
${s^\prime_{EM}}=\frac{s_{EM}}{k} \propto \frac{L_o}{k\cdot r^2_{EM}}$. This means that the new maximum distance corresponds to $r^\prime_{EM} = \sqrt{k}\cdot r_{EM}$, and thus 
$V^\prime_{EM} = k^{3/2}\cdot V_{EM}$.

An improvement of a factor of $k$ in the sensitivity of a GW detector yields 
${s^\prime_{GW}}=\frac{s_{GW}}{k} = \frac{L_o}{k\cdot r_{GW}}$. This corresponds to a new maximum distance of $r^\prime_{GW} =  {k}\cdot r_{GW}$, and in the observable volume of $V^\prime_{GW} = k^{3} \cdot V_{GW}$.
Numerically, if the sensitivity of an EM telescope is improved by a factor of (e.q.) three, then the number of observable objects of similar luminosity increases as a factor $N \propto 3^{3/2}\simeq 5$.
If the sensitivity of an interferometer for GWs is increased by a factor of three, the number of detectable sources increases by a factor of $3^3=27$.

\vskip 0.1cm \noindent $\bullet$ The frequencies of detectable GWs are below the few kHz range, and thus graviton energies $h\nu_{gw}$ are very small, making detection of  {individual quanta} extremely challenging (if not almost impossible in the near future).

\vskip 0.1cm \noindent $\bullet$ Gravitational radiation suffers a very small absorption when passing through ordinary matter. As a result, GWs can carry to us information about violent processes occurred in very dense environments. In the context of detection, in comparison even neutrinos have large scattering cross sections with matter. 

\vskip 0.1cm \noindent $\bullet$ It is almost impossible with current technology to detect man-made GWs. In a classic example from \cite{saulson}, if we consider a dumbbell consisting of two 1-ton compact masses with their centers separated by 2 m and spinning at 1 kHz (this is the limit for its stability), the strain $h$ to an observer 300 km away (in the far field) is $h \sim 10^{-38}$.

\subsection{The two-body system}
\label{sec:2body}

The quadrupole moment (\ref{eq:phgw14}) of a system of two point-like masses $m_1$ and $m_2$ in a binary orbit can be calculated. 
Here, a simple Newtonian approach is used that holds for low velocities. When the velocities become relativistic, the Newtonian framework used to derive relations between quantities no longer applies. One example is the case of the Kepler's third law 
\begin{equation}\label{eq:bygw8}
\omega^2_s = \frac{GM}{R^3} \quad \textrm{ where } M=m_1+m_2
\end{equation}
connecting angular velocity $\omega^2_s$ with the orbit size $R$. 
For small $R$ and high velocities (as the later stages of the inspiral, as discussed in the following) further \textit{post-Newtonian} approximations are necessary.
The post-Newtonian approximations are expansions terms of $v/c$ and are used for finding an approximate solution of the Einstein field equations for the metric tensor in the case of weak fields. 

\begin{figure}[tb]
\begin{center}
\includegraphics[width=8.0cm]{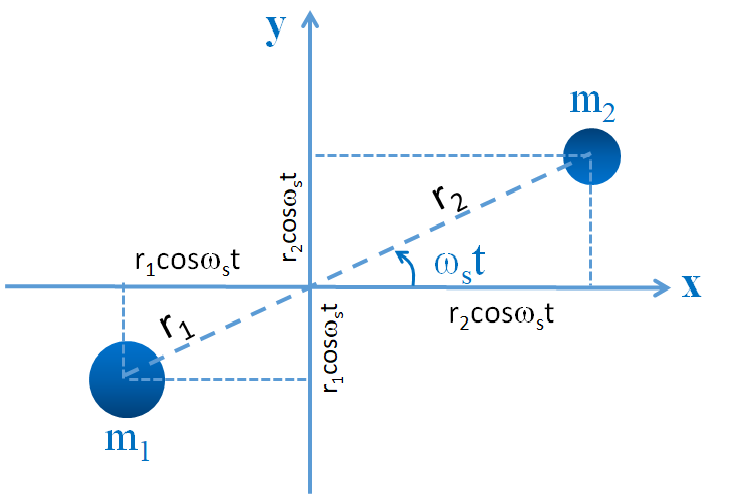}
\end{center}
\caption{\small\label{fig:binary} A two-body system, $m_1$ and $m_2$ orbiting in the $(x,y)$-plane around their center of mass.}
\end{figure}

We assume that the two-body system lies in the $(x,y)$-plane shown in Fig. \ref{fig:binary}; the quadrupole moment $Q_{ij}$ is computed using the Cartesian coordinate system $\textbf{x} = (x_1, x_2, x_3) = (x, y, z)$ whose origin is the center-of-mass; $r_i$ is the distance of the mass $i=1,2$ from the origin. Thus:
\begin{equation}\label{eq:bygw1}
Q_{ij} = \sum_{\alpha=1,2} m_\alpha 
{\begin{pmatrix}
\frac{2}{3}x_\alpha^2-\frac{1}{3}y_\alpha^2 & x_\alpha y_\alpha & 0\\
x_\alpha y_\alpha &\frac{2}{3} y^2_\alpha -\frac{1}{3} x^2_\alpha &0\\
0&0&-\frac{1}{3}r^2_\alpha\\
\end{pmatrix}} 
\end{equation}

In the simple case of a circular orbit at separation $R = r_1 + r_2$ and frequency $\nu_{s}$, angular velocity $\omega_s=2\pi\nu_{s}$ it is easy to derive with the help of Fig. \ref{fig:binary}: 
\begin{equation}\label{eq:bygw2}
Q^\alpha_{ij} = \frac{m_\alpha r^2_\alpha}{2} I_{ij}
\end{equation}
where the $3\times 3$ matrix $I$ is:
\begin{equation}
{\begin{pmatrix}
\cos(2\omega_s t) + \frac{1}{3} & \sin(2\omega_s t) & 0\\
\sin(2\omega_s t) & \frac{1}{3}-\cos(2\omega_s t)& 0\\
0 & 0 & - \frac{2}{3}\\
\end{pmatrix}} 
\end{equation}
By summing up the contribution of the two masses, we obtain:
\begin{equation}\label{eq:bygw3}
Q_{ij}= \sum_{\alpha=1,2} Q^\alpha_{ij} = \frac{1}{2} \mu R^2 J_{ij} \ ,
\end{equation}
where
\begin{equation}\label{eq:bygw4}
\mu \equiv \frac{m_1 m_2}{m_1+m_2}
\end{equation}
is the \textit{reduced mass} of the system. 

From Eq. (\ref{eq:phgw13}), the intensity $h_{ij}$ of the GW depends on the relative orientation of the observer with respect to the $(x,y)$-plane of the source. However, as given in (\ref{eq:phgw7}), in the direction perpendicular to the wave vector $\vec{k}$, there are only two degrees of freedom, expressed by the $h_+$ and $h_\times$ constant amplitudes.
To give a first-order estimate of the GW effect, let us assume that $I_{ij}=\cos(2\omega_s t)$ in (\ref{eq:bygw3}), i.e.  $ Q_{ij} \sim Q$ and that 
\begin{equation}\label{eq:bygw5}
h_\circ \sim h_+ \sim h_\times \ .
\end{equation}
Thus:
\begin{equation}\label{eq:bygw6}
\frac{d^2Q}{dt^2} = \frac{1}{2} \mu R^2 \cdot (4\omega_s^2) \cdot \cos(2\omega_s t) \ .
\end{equation}
The time-dependent wave amplitude is derived from the (\ref{eq:phgw13}):
\begin{equation}\label{eq:bygw7}
h(t) \simeq \frac{4G}{rc^4} \cdot (2 \mu R^2 \omega_s^2) \cdot \cos(2\omega_s t) = h_\circ \cos(\omega_{gw} t) \ ,
\end{equation}
where 
\begin{equation}
\omega_{gw} = 2 \omega_{s}.
\end{equation}
 
Notice that because the quadrupole moment is symmetric under rotations of an angle $\pi$ about the orbital axis, \textbf{the radiation has a frequency, $\nu_{gw}$, twice that of the orbital frequency of the source, $\nu_{s}$}.
Now, by using the Kepler's third law (\ref{eq:bygw8}), we can remove in (\ref{eq:bygw7}) the angular velocity and obtain:
\begin{equation}\label{eq:bygw9}
h_\circ \simeq \frac{4G}{rc^4} \cdot (2 \mu R^2)  \frac{GM}{R^3}
\end{equation}
or, equivalently:
\begin{equation}\label{eq:bygw10}
\boxed{
h_\circ= 2 \biggl( \frac{2GM}{c^2 r}\biggr) \biggl( \frac{2G\mu}{c^2 R}\biggr) =2 \frac{{\mathcal R}_{S_1}\cdot {\mathcal R}_{S_2}}{r\cdot R}
} \ .
\end{equation}
This is a relevant result: the strain $h_\circ$ derived from the quadrupole formula can be written into a manifestly dimensionless form by recognizing that the mass times $ 2G/c^2$ corresponds to the Schwarzschild radius ${\mathcal R}_{S_i}$ of the object (Eq. \ref{eq:rd1}).
At the denominator, the distance $R$ is an \textit{internal} parameter of the system, while $r$ is the distance of the source from the observer.
If the binary system consists of two neutron stars ($m_1 \simeq m_2 \simeq 1.4 M_\odot$), then both the Schwarzschild radii are $\sim$ 4 km.
If we consider two close-by neutron stars approaching their merging when $R\simeq 100$ km and at a distance of 40 Mpc from the Earth\footnote{1 parsec = $3.086\times 10^{16}$ m }, we obtain
\begin{equation}\label{eq:bygw11}
h_\circ= 2 \biggl( \frac{(4000 \textrm{ m})^2 }{10^5 \times 1.2\times 10^{24} \textrm{ m}^2}\biggr) \simeq 3\times 10^{-22} \ .
\end{equation}
Let us summarize the salient results in term of observable quantities. 
As a GW passes an observer, that observer will find spacetime distorted by the effects of strain. 
Distances $L$ between objects increase and decrease rhythmically as the wave passes, with a maximum amplitude $\Delta L_{max}$ such that
\begin{equation}\label{eq:bygw11b}
\frac{\Delta L_{max}}{L} \simeq h_\circ \ .
\end{equation}
with the pattern shown in Fig. \ref{fig:h+hx}, and at a frequency corresponding to that of the wave. 
To have a feeling, Eq. (\ref{eq:bygw11b}) means that the distance of the Earth from the Sun is changed by the distance of one atom during the passage of such GW. 
The frequency of the wave depends on the relative distance $R$ of the merging objects (in the Newtonian regime, according to the Kepler's third law (\ref{eq:bygw8})). The frequency interval 10 Hz-1000 Hz is particularly relevant. 
Thus, the quantity (\ref{eq:bygw11}) represents the order-of-magnitude of a detector sensibility to detect GW signals. 

Let us compute now the total luminosity (\ref{eq:enegw3}) of the source. 
The third derivative of (\ref{eq:bygw3}) yields the matrix:
\begin{equation}\label{eq:temp1}
{\dddot Q_{ij}}= \frac{1}{2} \mu R^2 (2\omega_s)^2 
{\begin{pmatrix}
\cos(2\omega_s t) & \sin(2\omega_s t) & 0\\
\sin(2\omega_s t) & -\cos(2\omega_s t)& 0\\
0 & 0 & 0 \\
\end{pmatrix}} 
\end{equation}
The double summation in (\ref{eq:enegw3}) yields a scalar (the sum of the product of the first line by the first column + second line times second column), explicitly:  
\begin{equation}\label{eq:temp2}
\sum_{i,j=1}^3 \frac{d^3 Q_{ij}}{dt^3} \frac{d^3 Q_{ij}}{dt^3}= 
[\cos^2(2\omega_s t) + \sin^2(2\omega_s t)] + 
[\sin^2(2\omega_s t) + \cos^2(2\omega_s t)] = 2
\end{equation}
Thus, the scalar quantity of Eq. (\ref{eq:enegw3}) becomes:
\begin{equation}\label{eq:enegw2}
{\mathcal L}= \frac{1}{5} \frac{G}{c^5}\cdot \biggl( \frac{1}{2} \mu R^2 \biggr)^2 \cdot (2\omega_s)^6 \cdot 2
        = \frac{32}{5} \frac{G}{c^5}\cdot [\mu R^2 \omega_s^3]^2
\end{equation}
In a similar way, the energy flux (\ref{eq:enegw1}) of a sinusoidal wave of angular frequency $\omega_{gw}$ and amplitude $h_\circ$ as obtained by using (\ref{eq:bygw7}) is:
\begin{equation}\label{eq:enegw1b}
{\mathcal F}= \frac{1}{32\pi} \frac{c^3}{G} h_\circ^2 \omega_{gw}^2  \ ,
\end{equation}
that for $\omega_{gw}=400$ s$^{-1}$ and $h_\circ=3\times 10^{-22}$ corresponds to 
${\mathcal F}= 7\times 10^{-5} \textrm{ W m}^{-2}= 0.07\ \textrm{ erg s}^{-1} \textrm{ cm}^{-2}$.
For comparison, typical fluxes measured by Fermi-LAT in the $\gamma$-ray band for steady sources are of the order of $10^{-11} \textrm{ erg s}^{-1} \textrm{ cm}^{-2}$. Hence, during the time interval $\Delta t\sim 1/\nu_{gw}$ when the waves of a coalescing binary neutron star system 40 Mpc away pass the Earth, the energy flux is order of $10^{10}$ that for a steady source of $\gamma$-rays. However, as shown below, detecting the passage of this energy flux is a formidable experimental challenge.

\section{Ground-based laser interferometers}
\label{sec:experiments}

To enable sensitivity to a wide range of astrophysical GW sources, ground-based interferometers must thus be designed to achieve strain down to $\sim 10^{-22}$, or better, possibly over the widest frequency range in the 10-5000 Hz
\footnote{As the standard range of audible frequencies is 20 to 20,000 Hz, the signal of the passage of a GW can be transduced to a sound audible by human ears. There are different examples on the educational resources webpages of the experiments, \url{https://www.ligo.caltech.edu/}. However, remember that this is just a didactic and sociological trick and GWs are not detected by acoustic devices.}.

\begin{figure}[tb]
\begin{center}
\includegraphics[width=12.0cm]{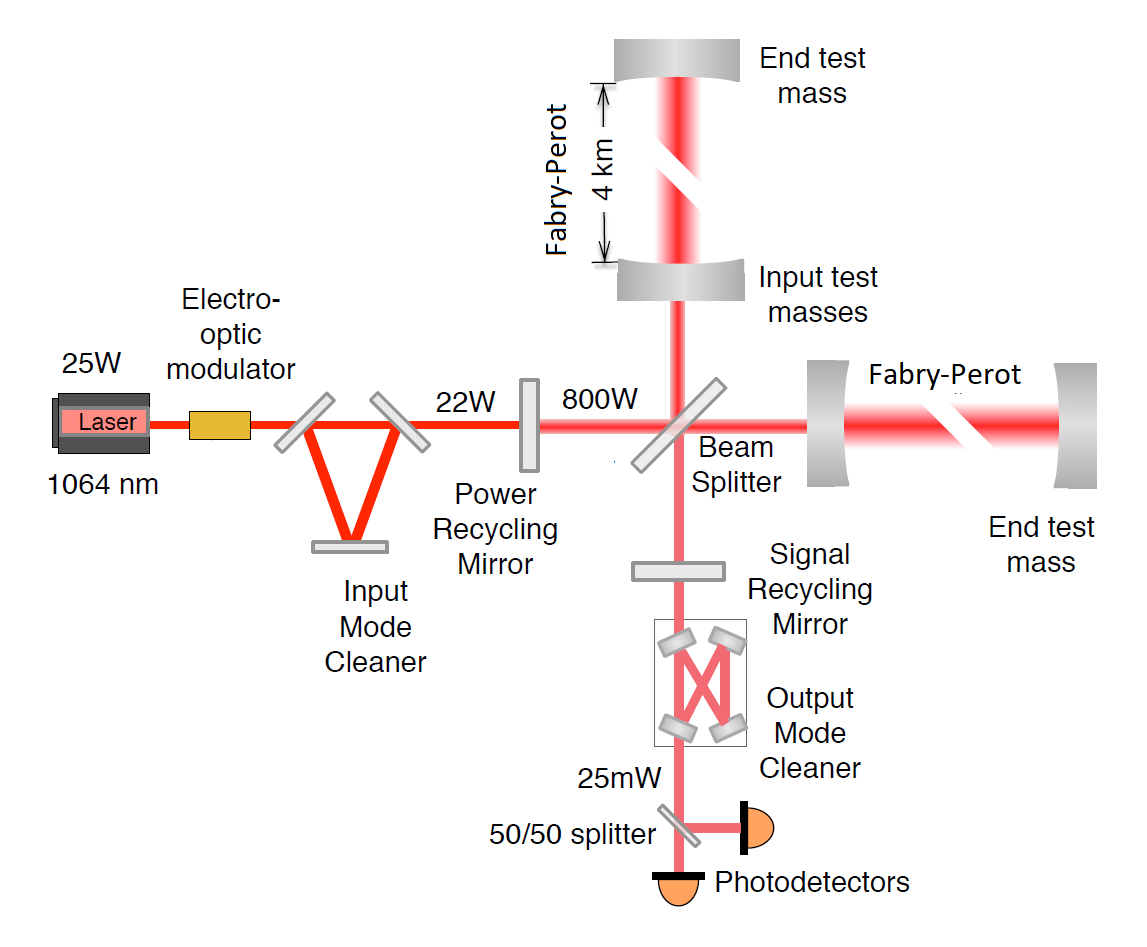}
\end{center}
\caption{\small\label{fig:LIGO} Layout of an aLIGO detector. 
Adapted from \cite{tec}. See text for details.}
\end{figure}

Ground-based interferometers are arranged in the Michelson configuration (L-shaped). They consists of a laser, a beam splitter, a series of mirrors and photodetectors that records the interference pattern, see Fig. \ref{fig:LIGO}.
The laser beam passes through a \textit{beam splitter} that splits a single beam into two identical beams, one of which at 90$^\circ$. 
Each beam then travels down an arm of the interferometer. At the end of each arm, a mirror acting as \textit{test mass} reflects each beam back to the beam splitter where the two beams merge back into a single beam. 
In 'merging', the light waves from the two beams interfere with each other before reaching a \textit{photodetector}. 
GW interferometers are set up so that the interference is destructive at the photodetector. Any change in light \textbf{intensity} due to a different interference pattern indicates that something (noise or signal) happened to change the distance $L$ traveled by one or both laser beams. Moreover, the interference pattern can be used to calculate precisely $\Delta L/L$, i.e. the signal strain (\ref{eq:bygw11b}).
This point is of fundamental importance: \textbf{the interferometer is sensible to the phase of the quantity $\mathbf{h(t)}$ (the strain, as that given in Eq. (\ref{eq:bygw7})), and not to the GW energy flux, Eq. (\ref{eq:enegw1}). The former decreases as $\mathbf{1/r}$, the latter as $\mathbf{1/r^2}$. }

A GW observatory cannot operate alone. A coincident detection with two interferometers reduces the noise background and improves the possibility of the source localization. These objectives are even more improved when interferometers are connected in a network, as in the present configuration of GW observatories. 

\begin{figure}[tb]
\begin{center}
\includegraphics[width=12.0cm]{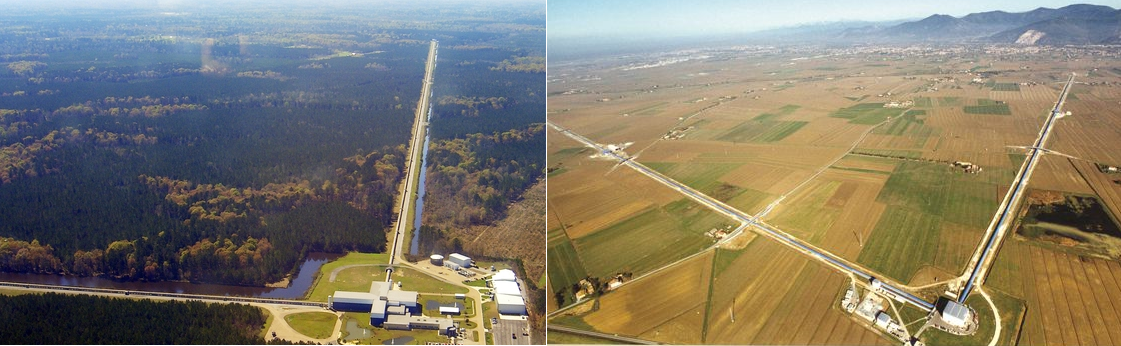}
\end{center}
\caption{\small\label{fig:LigoVirgo} 
Left: Aerial view of the LIGO gravitational wave detector in Livingston, Louisiana. (Credit: Caltech/MIT/LIGO Lab). A similar detector exists in the Washington State (LIGO Hanford). Right: Aerial view of the Virgo gravitational wave interferometer in Italy (Credit: EGO/Virgo)
}
\end{figure}

LIGO consists of two widely separated (about 3000 km) identical detector sites in USA working as a single observatory: one in southeastern Washington State and the other in rural Livingston, Louisiana, Fig. \ref{fig:LigoVirgo} left.  
The LIGO Scientific Collaboration (LSC) includes scientists from both LIGO laboratories and collaborating institutions. LSC members have access to the GEO 600 detector in Germany.
Virgo is a 3 km interferometer located outside of Pisa, Italy, funded by the European Gravitational Observatory (EGO), a collaboration between the Italian INFN and the French CNRS, Fig. \ref{fig:LigoVirgo} right.  
While the LSC and the Virgo Collaboration are separate organizations, they cooperate closely; they are referred to as LVC, and they sign collectively the research papers.

Initial LIGO (iLIGO) took data between 2001 and 2010, almost contemporary with initial Virgo, without detecting GWs. 
The redesign, construction, preparation and installation of the Advanced LIGO (aLIGO) took 7 years (from 2008 to 2015), and for the Advanced Virgo from 2010 to 2017.
The improvements had the objective of making the observatories 10 times more sensitive, allowing to increase the volume of the observable universe by a factor of 1000.
In September 2015, aLIGO began the era of GW astronomy with its first observation run (O1) and detections, collecting data until January 2016. The interferometers were not yet operating at design sensitivity during O1.
The second observing run (O2) of aLIGO started on November 30, 2016. aVirgo joined the O2 run on August 1, 2017. Both ended O2 operations on August 25, 2017. 

\begin{figure}[tb]
\begin{center}
\includegraphics[width=10.0cm]{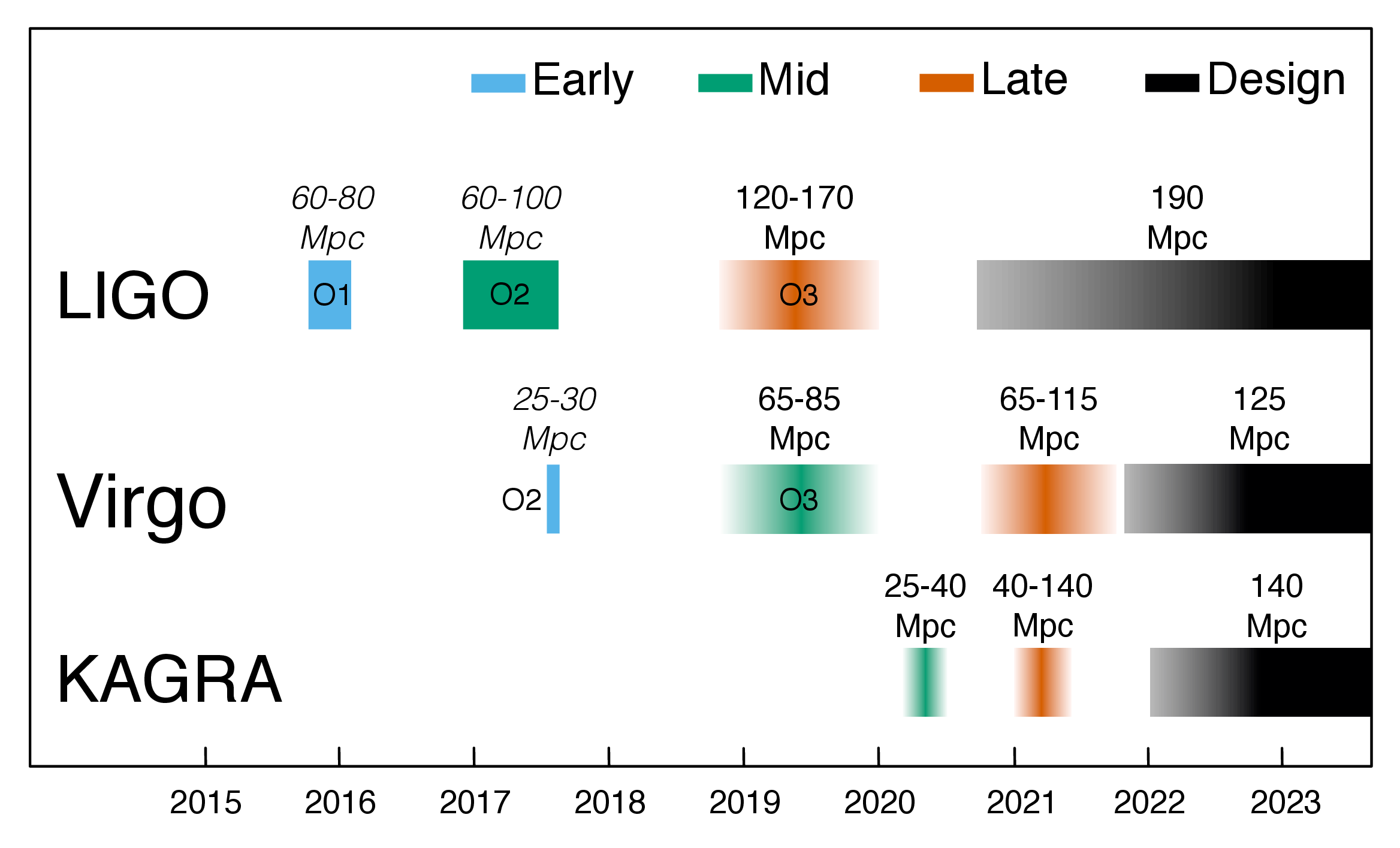}
\end{center}
\caption{\small\label{fig:LVKtime} 
Indicative timeline for observing runs (colored bars) with the LIGO, Virgo and KAGRA detectors over the coming years. Between observing runs, the detectors are tuned to improve sensitivities. The numbers above the bars correspond to the average distance (in Mpc) within which a binary neutron star merging can be observed. (Credit: LVC) }
\end{figure}

The O3 observing run started with both LIGO detectors and Virgo on April 2019 and it will last for about 12 months. Looking forward, the observing plan includes the Japanese detector KAGRA in 2020. The two LIGO detectors, Virgo, and KAGRA should all reach the planned optimal sensitivities by 2022. A further detector, LIGO-India, will also be added. The increasing number of detectors in the network increases the observation duty cycle and makes it easier to detect signals and helps in the source localization. Fig. \ref{fig:LVKtime} shows a plausible timeline for observing with the LIGO, Virgo and KAGRA detectors. 

In the following, some details of the design of the interferometers are described, referring in particular to Fig. \ref{fig:LIGO}. The most impressive technology resides in their laser, seismic isolation systems necessary to remove unwanted vibrations, vacuum systems, optics components to preserve laser light and power, and computing infrastructure to handle in real time collected data.
Some quantities (as the number of reflections, laser power, etc.) slightly change from run O1, O2 and final design.   
We specialize the description to the aLIGO setup; the Virgo interferometer works similarly. 

\vskip 0.2cm
\noindent  \textbf{The optics system}
of GW interferometers consists of lasers, a series of mirrors, and photodetectors.
If LIGO's interferometers were basic Michelson's, even with arms 4 km long, they would still not be long enough to be sensible to GWs.
Fundamental tools are \textit{Fabry-Perot cavities}.
A Fabry-Perot cavity is created by adding mirrors near the \textit{beam splitter} that continually reflect parts of each laser beam back and forth within the long arms. In aLIGO, this occur about 270 times before the laser beams are merged together again, making LIGO's interferometer arms effectively 1080 km long.

A second design factor important to improve the interferometer's resolution is the laser power. The more photons that merge at the beam splitter, the sharper the resulting interference pattern becomes.
To reach the sensitivity necessary for a discovery ($h\sim 10^{-22}$), the laser must reach a much higher power (see the discussion in Appendix \ref{sec:sensitivity}).
For this reason, an additional device, the \textit{power recycling mirrors} are placed between the laser source and the beam splitter to boost the power of LIGO laser: in O1 run this power is increased by a factor of $\sim 40$. 
Similarly to the beam splitter itself, the power recycling mirror is only partly reflective and the light from the laser first passes through the mirror to reach the beam splitter. 
The instrument is accurately aligned in a way that the largest fraction of the reflected laser light from the arms follows a path back to the recycling mirrors rather than to the photodetector. 
These 'recycled' photons add to the ones just entering. 
As a further difference with simple Michelson interferometers, aLIGO possess \textit{signal recycling} mirrors, which like power recycling, enhance the output signal.

Before entering the power recycling mirror, the \textit{input mode cleaner} is a suspended, triangular Fabry-Perot cavity needed to clean up the spatial profile of the laser beam, clean polarization, and help stabilize the laser frequency. Similarly, before the photodetector, an \textit{output mode cleaner} is present at the antisymmetric port, to reject unwanted spatial and frequency components of the light, before the signal is detected.

\vskip 0.2cm
\noindent  \textbf{The laser.}
The heart of LIGO is its Nd:YAG laser, with wavelength $\lambda=1064$ nm. The maximum power is $\sim$ 200 W by design, but only 22W were used in run O1. It takes different steps to amplify its power and refine its wavelength to the level necessary for the experiment.
The first step is a laser diode generating an 808 nm near-infrared beam of $\sim$ 4 W (about 800 times more powerful than standard laser pointers).
Then, the 4 W beam enters a device consisting of a small boat-shaped crystal and it bounces around inside this crystal and stimulates the emission of a 2 W beam with a wavelength of 1064 nm, in the invisible infrared part of the spectrum. Another amplifying device boosts the 1064 nm beam from 2 to 35 W. 
Finally, a High Powered Oscillator performs further amplification and refinement, and generates the final beam. 

\vskip 0.2cm
\noindent \textbf{Mirrors.}
The suspended primary mirrors act as the \textit{test masses}, and must be of the highest quality available, both in material and shape.
LIGO's mirrors weigh 40 kg each and are made of very pure fused silica glass. The mirrors were polished so precisely that the difference between the theoretical design and the actual polished surface is measured in atoms. 
They reflect most of the laser light and absorb just one in $3\times 10^6$ hitting photons, avoiding the mirror heating. 
The heating could alter the mirror shapes enough that they degrade the quality of the laser light. 
The mirrors also refocus the laser, keeping the beam traveling coherently throughout its multiple reflections before arriving at the photodetector.

\vskip 0.2cm
\noindent \textbf{Seismic isolation.}
Laser interferometers are extremely sensitive to all vibrations near (such as trucks driving on nearby roads) and far (earthquakes, nearby and far away).
The suspended primary mirrors must be as free as possible, i.e. decoupled from any man-made or earthly vibrations. 
For this reason, \textit{active} and \textit{passive} damping systems are used to eliminate vibrations.
The active damping consists of a system of sensors designed to feel different frequencies of ground movements. 
These sensors work side-by-side and send their feedbacks to a computer that generates a net counter-motion to cancel all of the vibrations simultaneously. 
The passive damping system holds all test masses (its mirrors) perfectly still through a 4-stage pendulum called a \textit{quad}. 
At the end of the quad, LIGO's mirrors are suspended by 0.4 mm thick fused-silica (glass) fibers. 
The configuration absorbs any movement not completely canceled out by the active system.

\vskip 0.2cm
\noindent \textbf{Vacuum.}
The laser beam travel in one of the largest and purest sustained vacuums on Earth ($10^{-6}$ Pa). 
The presence of dust into the path of the laser, or worse, onto a mirror can cause some of the light to scatter (i.e., be reflected in some random direction away from its path).
The presence of air produces an index of refraction that could affect the apparent distance between the mirrors. In addition, molecules of air hitting the mirrors due to the Brownian motion can cause them to move, masking the signal strain. 
Many techniques are used to remove all the air and other molecules from vacuum tubes; for instance, the tubes were heated (between 150 C and 170 C) for 30 days to drive out residual gas molecules and turbo-pumps sucked out the bulk of the air contained in the tubes. Finally, ion pumps operating continuously maintain the vacuum by extracting individual remaining gas molecules.
It took about 40 days to remove $\sim 10^4$ m$^3$ of air and other residual gases from each of vacuum tubes, before starting of the physics runs.  

\vskip 0.2cm
\noindent \textbf{Computation and Data Collection.}
Computers are required both to run the LIGO instruments and to process the data that it collects.
When it is in \textit{observing} mode, an interferometer generates TB of data every day that must be transferred to a network of supercomputers for storage and archiving. 
Because much of the astrophysical information are extracted from the phase of the GW, different kinds of data analysis methods are employed than the ones normally used in astronomy. They are based on matched filtering and searches over large parameter spaces of potential signals. This style of data analysis requires the input of pre-calculated template signals, which means that GW detection depends more strongly than most other branches of astronomy on theoretical input modeled at computer.

\section{GW150914}\label{sec:gw15}

\textit{``On September 14, 2015 at 09:50:45 UTC the two detectors of the Laser Interferometer Gravitational-Wave Observatory simultaneously observed a transient gravitational-wave signal. The signal sweeps upwards in frequency from 35 to 250 Hz with a peak gravitational-wave strain of $1.0\times 10^{-21}$. It matches the waveform predicted by general relativity for the inspiral and merger of a pair of black holes and the ringdown of the resulting single black hole. 
The signal was observed with a matched-filter signal-to-noise ratio of 24 and a false alarm rate estimated to be less than 1 event per 203 000 years, equivalent to a significance greater than 5.1 ${\sigma}$. 
The source lies at a luminosity distance of $410^{+160}_{-180}$ Mpc corresponding to a redshift $z=0.09^{+0.03}_{-0.04}$. 
In the source frame, the initial black hole masses are $36^{+5}_{-4} M_\odot$ and $29^{+4}_{-4} M_\odot$, and the final black hole mass is $62^{+4}_{-4} M_\odot$, with $3.0^{+0.5}_{-0.5} M_\odot c^2$ radiated in gravitational waves. All uncertainties define 90\% credible intervals. These observations demonstrate the existence of binary stellar-mass black hole systems. This is the first direct detection of gravitational waves and the first observation of a binary black hole merger. ''}

The text reproduced above is the abstract of one of most important papers in the history of science \cite{1}, opening the field of astrophysics with gravitational waves. 

The theoretical work started in the 1970's led to the understanding of GWs produced by the merging of two BHs through the so-called ``quasinormal'' emission. Mathematically, the solutions of the Einstein equations foreseen complex frequencies, with the real part representing the actual frequency of the oscillation and the imaginary part representing a damping. 
In the 1990's higher-order post-Newtonian calculations preceded extensive analytical studies. 
These improvements, together with the significant contribution of \textit{numerical relativity}, have enabled modeling of binary BH mergers and accurate predictions of their gravitational waveforms. 

\begin{figure}[tb]
\begin{center}
\includegraphics[width=11.0cm]{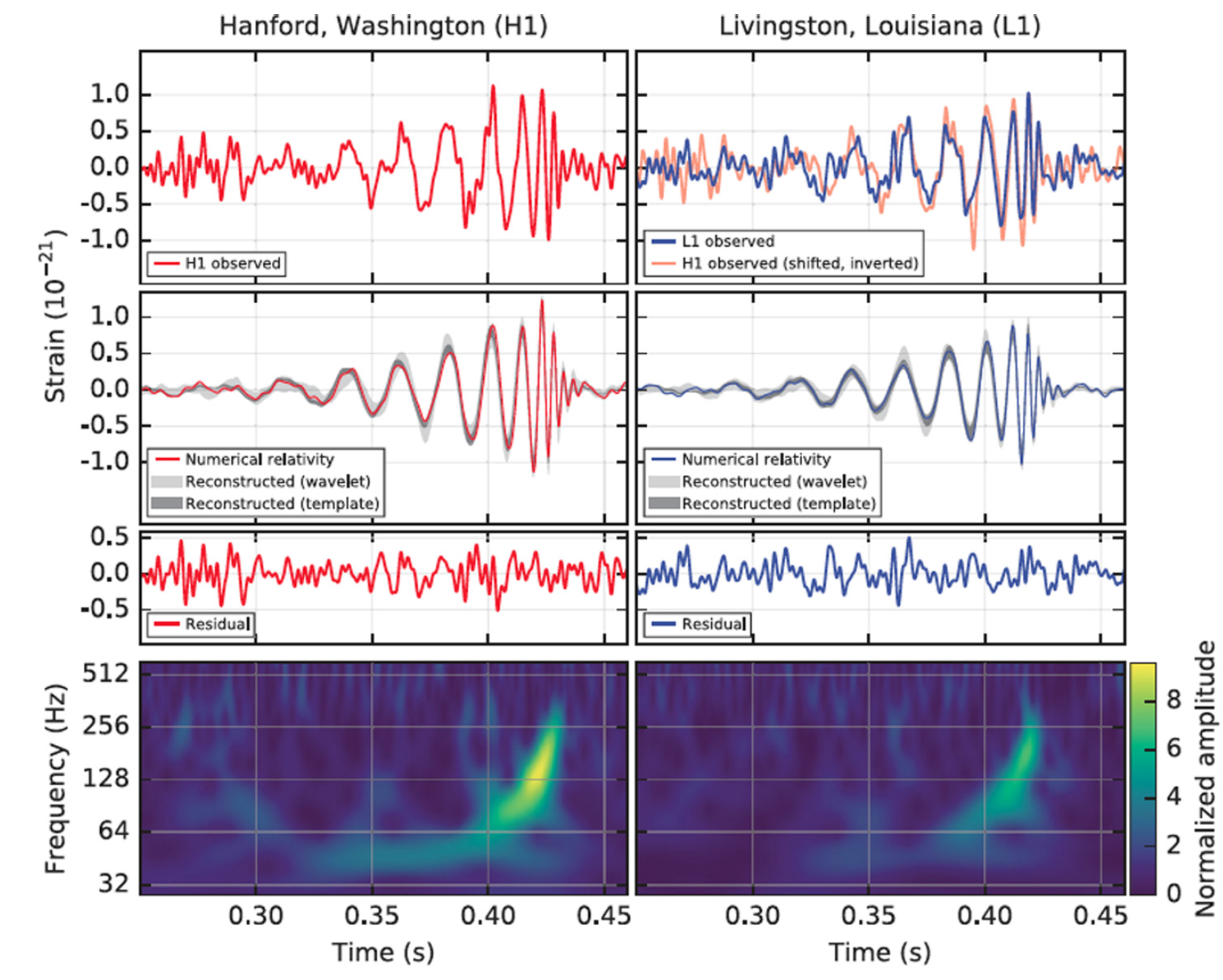}
\end{center}
\caption{\small\label{fig:GW150914} Summary of LIGO data (Fig. 1 of \cite{1}) for GW150914. The top left (right) panel shows the strain $h$ observed by the Hanford (Livingston) detector as a function of time. Spectral noise features in the detectors have been filtered. The second row shows a fit to the data using sine-Gaussian wavelets (light gray) and a different waveform reconstruction (dark gray). Also shown in color are the signals obtained from numerical relativity using the best-fit parameters to the data. The third row shows the residuals obtained by subtracting the numerical relativity curve from the filtered data in the first row. The fourth row gives a time-frequency representation of the data and shows the signal frequency increasing over time (the \textit{chirp effect). }}
\end{figure}

Binary BH mergers take place in three stages, as evident in Fig. \ref{fig:GW150914} and drawn on top of Fig. \ref{fig:chirp}. 
Initially, they circle their common center of mass in essentially circular orbits (\textit{inspiral}). They lose orbital energy in the form of gravitational radiation and they spiral inward. In the second stage (\textit{merging}), the two objects coalesce to form a single BH. In the third stage (\textit{ringdown}), the merged object relaxes into its equilibrium state, a Kerr black hole. 
The LIGO/Virgo collaboration for the search of a GW signal in the data stream make use of a formalism that defines many templates of \textit{matched-filter signal to noise ratio} combining results from the post-Newtonian approach with results from perturbation theory and numerical relativity. 
In particular, GW emission from binary systems with $h\gtrsim 10^{-22}$, individual masses from 1 to 99 $M_\odot$, and dimensionless spins (see \S \ref{sec:spin}) up to $\chi=0.99$ were searched for. 
For GW150914, approximately 250,000 template waveforms have been used to cover the parameter space.

We shall try to derive, by inspection of the detector data reported in Fig. \ref{fig:GW150914} and the physics of GWs produced by binary systems described in the previous sections, the main results described in the abstract.

\subsection{Inspiral stage}

The initial {inspiral} phase occurs when the BHs rotate non-relativistically their common center of mass in circular orbits, as in Fig. \ref{fig:binary}. 
Thus, Newtonian mechanics apply and the angular frequency $\omega_s$ is related to the separation of the two black holes, $R$, via Kepler's third law (\ref{eq:bygw8}).

Let consider now the \textit{orbital energy} and its variation with time.
The total energy is the sum of the kinetic, $K$, and potential, $U$, energies. In the gravitationally bound system of Fig. \ref{fig:binary} we have
\begin{equation}\label{eq:ene1}
E_{tot}= K+U= \frac{1}{2} \mu \omega_s^2 R^2 - \frac{Gm_1m_2}{R}
= -\frac{GM\mu}{2R}=   -\frac{G m_1 m_2}{2R} \ . 
\end{equation}
This is a well-known equation (it corresponds to the \textit{virial theorem} in multi-body systems); in our case, it gives the total energy of the system as a function of the BH separation.

Classically, there is no gravitational radiation and the circular orbit will persist forever. In general relativity, the orbiting BHs will emit gravitational radiation thereby losing energy and spiraling towards each other\footnote{Circular orbits are used for simplicity, but careful analysis shows that even if the orbits were initially elliptical then emission of GWs will quickly produce circular orbits.}.
At large distance/low $\omega_{s}$ it is easy to see from (\ref{eq:bygw7}) that $h_\circ$ is small and not measurable in a detector.  As the BHs lose orbital energy in the form of gravitational radiation, they spiral inward. 
If the radius of the orbit decreases, also the total energy (\ref{eq:ene1}) decreases at a rate
\begin{equation}\label{eq:ene2}
\frac{dE_{tot}}{dt}= \frac{GM\mu}{2R^2}\frac{dR}{dt}
=\frac{GM\mu}{2R} \frac{\dot R}{R} 
\end{equation}
that must be numerically equal to the power emitted as gravitational radiation, Eq. (\ref{eq:enegw2}). 
According to the Kepler's third law, also the angular velocity changes, by increasing in time, as obtained by differentiation of (\ref{eq:bygw8}):
\begin{equation}\label{eq:ene3}
\frac{\dot \omega_s}{\omega_s} = -\frac{3}{2} \frac{\dot R}{R} .
\end{equation}

If we want to know the mass of the system that produce the wave, we must correlate to the observables in Fig. \ref{fig:binary}, namely: the measured strain $h$, the frequency of the wave $\nu_{gw}$, and its derivative, $\dot \nu_{gw}$.
Thus:
\begin{equation}\label{eq:ene4}
\frac{dE_{gw}}{dt}= -\frac{dE_{tot}}{dt}= 
-\frac{GM\mu}{2R} \frac{\dot R}{R}= 
\frac{GM\mu}{3R}\frac{\dot \omega_s}{\omega_s} \ .
\end{equation}
The left-hand side of this equation can be replaced with the energy flux of the gravitational wave obtained in (\ref{eq:enegw2}):
\begin{equation}\label{eq:ene5}
\frac{32}{5}\frac{G}{c^5} \mu^2 R^4 \omega_s^6 = \frac{GM\mu}{3R}\frac{\dot \omega_s}{\omega_s}
\end{equation}
that numerically depends on the masses, on the radius $R$, on the frequency and its time derivative. 
We can make $\dot \omega_s$ explicit in (\ref{eq:ene5}):
\begin{equation}\label{eq:ene6}
{\dot \omega_s} =\frac{96}{5}
\frac{\mu \omega_s^7 ({GM}/{\omega_s^2})^{5/3} } {Mc^5 } \ ,
\end{equation}
where we removed the $R^5$ term using the Kepler's third law. 
\begin{figure}[tbh]
\begin{center}
\includegraphics[width=9.5cm]{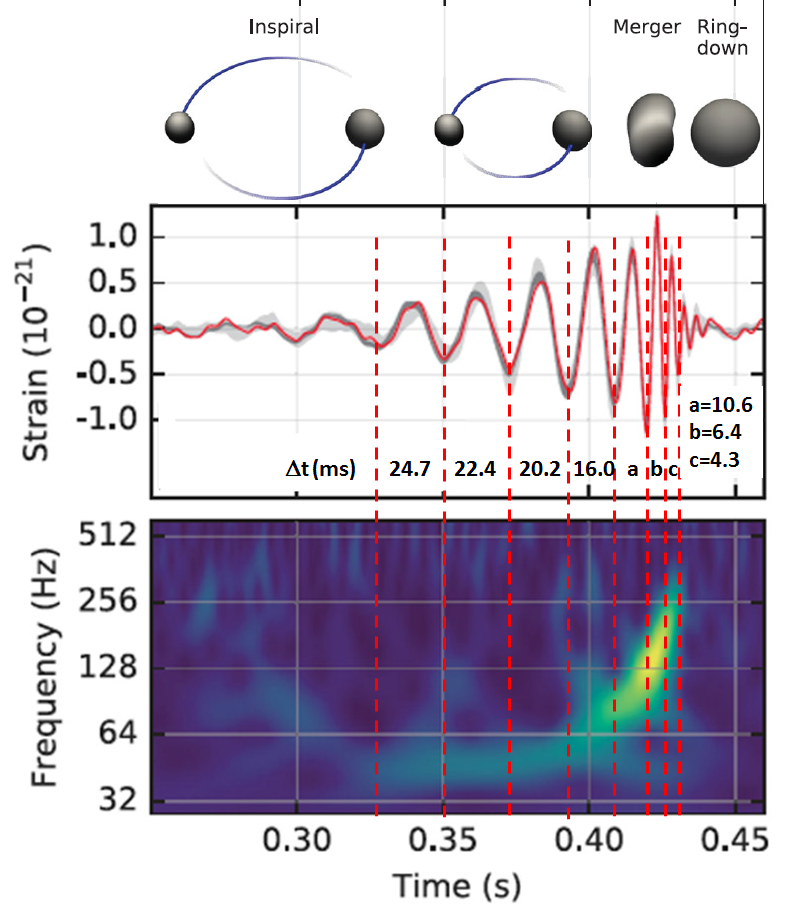}
\end{center}
\vspace{-5mm}
\caption{\small\label{fig:chirp} Detail of Fig. \ref{fig:GW150914} with the strain $h$ observed by the Hanford detector. The vertical dashed line are used to compute the time interval $\Delta t$ (ms) between two successive minima; the values are reported in Tab. \ref{tab:chirp} and used to obtain $\nu_{gw}$ and $\dot \nu_{gw}$ and thus the chirp mass. On top, a sketch of the three stages of the event.
} 
\end{figure}
This equation can be rewritten as
\begin{equation}\label{eq:ene7}
{\dot \omega_s^3} = \biggl(\frac{96}{5}\biggr)^3
\frac{G^5}{c^{15}} \mu^3 M^2 \omega_s^{11} 
= \biggl(\frac{96}{5}\biggr)^3
\frac{\omega_s^{11}}{c^{15}} \cdot (G {\mathcal M})^5
\end{equation}
where the so-called \textit{chirp mass} ${\mathcal M}$, is defined as:
\begin{equation}\label{eq:ene8}
{\mathcal M} \equiv (\mu^3 M^2)^{1/5} =  
\frac{(m_1 m_2)^{3/5}} {(m_1+m_2)^{1/5}} \ .
\end{equation}
The value of the chirp mass is a crucial scale in the inspiral process, and it can be derived by inverting (\ref{eq:ene7}):
\begin{equation}\label{eq:ene9}
{\mathcal M} = \frac{c^3}{G} \biggl[ \biggl(\frac{5}{96}\biggr)^3    \omega_s^{-11}\ {\dot \omega_s}^3\biggr]^{1/5}
\ .
\end{equation}
In order to obtain the chirp mass from data, it helps to rewrite Eq. (\ref{eq:ene9}) in terms of the frequency $\nu_{gw}$ of the \textit{observed} radiation. 
Remembering that $\omega_{gw}=2\omega_s$, Eq. (\ref{eq:bygw7}), thus $\pi \nu_{gw}=\omega_s$. Making this substitution in Eq. (\ref{eq:ene9}) we obtain
\begin{equation}\label{eq:ene10}
{\mathcal M} = \frac{c^3}{G}  \biggl(\frac{5}{96} \pi^{-8/3}\ \nu_{gw}^{-11/3}\ {\dot \nu_{gw}} \biggr)^{3/5}
\ ,
\end{equation}
that precisely matches the only equation in \cite{1}.
Equation (\ref{eq:ene10}) shows that as the BHs spiral inward, the frequency of the GW increases rapidly. This is the famous \textit{chirp} effect, visible in the bottom panel of Fig. \ref{fig:GW150914}.

We can compute the chirp mass $\mathcal M$ by extracting the values of time $\Delta t$ between successive minima in the strain $h$ of from Fig. \ref{fig:chirp}, and reported in the first column of Table \ref{tab:chirp}. 
Then, ${\nu_{gw}}=1/\Delta t$ and ${\dot \nu_{gw}}= \Delta \nu_{gw}/\Delta t $ are reported in the second and third columns. 
According to (\ref{eq:ene10}), the product $\nu_{gw}^{11}\ \dot \nu_{gw}^3$ (forth column) must be constant and connected with the value of the chirp mass $\mathcal M$ at different phases (fifth column).
Thus, the characteristic mass scale ${\mathcal M} \simeq 30 M_\odot $
of the radiating system is obtained by direct inspection of the time-frequency behavior of data, in agreement with the value reported in \cite{1}.
The last column contains the distance $R$ between BHs during the different cycles reported in the figure. It can be noticed that $R$ is incredibly small with respect to normal length scales for stars. 
\begin{table}[t]
\caption{\small The first column reports the value of the $\Delta t$ between successive minimum as obtained from Fig. \ref{fig:chirp}, and the second column the frequency change rate of the gravitational wave, $\dot \nu_{gw}$, evaluated as $\Delta \nu_{gw}/\Delta t$. In the following columns: $\nu_{gw}^{11}\ \dot \nu_{gw}^3$; the chirp mass  ${\mathcal M}$, Eq. (\ref{eq:ene10}), and its ratio with the solar mass, $M_\odot$. Finally, in the last column, the distance $R$ between the two objects evaluated with the Kepler's third law (remember: $\omega_s = \pi \nu_{gw}$) }
\label{tab:chirp}
\begin{center}
\begin{tabular}{ccccccc}
\hline
$\Delta t$ & $\nu_{gw}$ & $\dot \nu_{gw}$ & $\nu_{gw}^{11}\ \dot \nu_{gw}^3$  & ${\mathcal M}$ & ${\mathcal M}/M_\odot$ & R \\
(ms)       & (Hz) & (Hz s$^{-1})$ & & (kg) & & (km) \\
\hline
24.7 & 40 & - & - & - & - & 630 \\				
22.4 &45 &	186 &	4.6$\times 10^{-12}$ &	6.0$\times 10^{+31}$ &	30 & 590 \\
20.2 &50 &	241 &	3.2$\times 10^{-12}$ &	5.6$\times 10^{+31}$ &	28 & 550 \\
16.0 & 63 & 812 &	9.4$\times 10^{-12}$ &	7.0$\times 10^{+31}$ &	35 & 470 \\
10.6	& 94 & 3004 &	5.1$\times 10^{-12}$ &	6.2$\times 10^{+31}$ &	31 & 360 \\
6.4	& 156	& 9673 &	6.7$\times 10^{-13}$ &	4.1$\times 10^{+31}$ &	21 & 255 \\
4.3	& 233	& 17746 &	5.2$\times 10^{-14}$ &	2.5$\times 10^{+31}$ &	12 & 200 \\
\hline
\end{tabular}
\end{center}
\end{table} 

The chirp mass is a quantity that depends on the two BH masses but, by itself, it does not reveal their individual values.
 For identical objects ($m_1=m_2$, likely condition for a system of two NSs), then the total mass $M=m_1+m_2$ corresponds to $M= 4^{3/5}{\mathcal M}\simeq 2.3 {\mathcal M}$. 
More generally, the total mass of the pair has to be greater than $4^{3/5}{\mathcal M}$. In fact, if
\begin{equation}\label{eq:mas1}
m_1=\alpha M  \quad ; \quad m_2=(1-\alpha) M
\end{equation}
then, from the definition (\ref{eq:ene8}):
\begin{equation}\label{eq:mas2}
M=\frac{{\mathcal M}}{[(\alpha (1-\alpha)]^{3/5}} \ .
\end{equation}
The denominator is maximum for $\alpha=1/2$, and thus $M$ is $minimum$ for a system with equal masses. If the two BHs in GW150914 are equal, then the minimum total mass of the system is $M=2.3 {\mathcal M}\simeq 69\ M_\odot$.

When the two BHs approaches, the values of ${\mathcal M}$ in the last two rows of Table \ref{tab:chirp} significantly deviates from previous values: the validity of the Newtonian approach does not hold anymore, and also spin effects start to be significant. 
The observables in the second stage can be used to derive the values of the two individual masses. 

\vskip 0.3cm
\noindent\textbf{Exercise:} Estimate the speed of the masses in Table \ref{tab:chirp}). 
\vskip 0.3cm

\subsection{Coalescence stage: individual masses} 

In the second stage of the recorded signal of GW150914, both the frequency and the strain increase, and the BHs {coalesce} to form a single BH. 

The gravitational radiation emitted during the inspiral stage can be described with the simple Newtonian approach; as the distance between objects decreases and angular velocity increases, the radiation luminosity increases,  see Eq. (\ref{eq:enegw2}). 
Thus, the computation of observables during the merger is less simple than in the inspiral stage. 
The merger presents a formidable problem that has be faced on only recently with numerical relativity.

A rational choice for the beginning of the coalescence is the moment when the separation of the two BHs is equal to the sum of their Schwarzschild radii. This can be expressed, using (\ref{eq:rd1}), as:
\begin{equation}\label{eq:co1}
{\mathcal R} = \frac{2G}{c^2}(m_1+m_2)
\end{equation}
For $M= m_1+m_2 \simeq 70 M_\odot$, the corresponding Schwarzschild radius is ${\mathcal R}\simeq 200$ km. This agrees with the minimum observable distance reported on Table \ref{tab:chirp}.
At this value of ${\mathcal R}$ and $M=m_1+m_2$ corresponds, from the Kepler's third law (\ref{eq:bygw8}), an angular velocity of
\begin{equation}\label{eq:co2}
\omega_{Schw} = \frac{1}{\sqrt 8} \frac{c^3}{GM} \ .
\end{equation}
From inspection of the bottom panel of Fig. \ref{fig:chirp}, a signal is visible up to (roughly) the half of the bin between 256 and 512 Hz. This corresponds (because the non-linear scale) to a maximum visible frequency of the gravitational wave of 
\begin{equation}\label{eq:co3}
\nu_{gw}^{max} \simeq 330 \textrm{ Hz} \ .
\end{equation}
By inverting (\ref{eq:co2}) and using the maximum observable frequency to estimate $\omega_{Schw} \simeq \pi \nu_{gw}^{max}$ (remember always the factor of two between the frequency of the wave and that of the system) we obtain:
\begin{equation}\label{eq:co4}
M= \frac{1}{\pi \sqrt 8}\frac{c^3}{G\nu_{gw}^{max}}
 = 1.38\times 10^{32} \textrm{ kg} \simeq 70 M_\odot \ ,
\end{equation}
a value close to the minimum $M$. 
Thus, we have obtained from inspection of the data at the detector on Earth (and for this reason, we add now a superscript to the values)
${\mathcal M}^{det} = 30 M_\odot$ and $M^{det}=70 M_\odot$.  
Those two value can be used to determine the individual masses of the BHs. Using (\ref{eq:mas2}), we derive a value of $\alpha \simeq 0.6$ and thus:
\begin{equation}\label{eq:co5}
m^{det}_1 =\alpha M^{det} = 42 M_\odot  \quad ; \quad m^{det}_2 =(1-\alpha) M^{det} = 28 M_\odot  \ .
\end{equation}
After the correction for cosmological effects (next subsection), these values are compatible, within errors, with that obtained from the LIGO/Virgo Collaboration and reported on the abstract of the paper.

\subsection{Luminosity Distance and Cosmological Effects \label{sec:cosmo}}

An estimate of the distance of the system can be obtained through the relation between the intrinsic and observed luminosity.
The \textit{luminosity distance} $D_L$ is defined in terms of the relationship between the effective luminosity of the object, ${\mathcal L}$, and its energy flux, ${\mathcal F}$:
\begin{equation}\label{eq:dl1}
{\mathcal F} = \frac{{\mathcal L}}{4\pi D_L^2}
\end{equation}
where ${\mathcal F}$ is the flux (W m$^{-2})$, and ${\mathcal L}$  is the luminosity (W).
Neglecting, as a first approximation for GW150914, cosmological corrections (to be verified a posteriori), and using ${\mathcal F}$ from (\ref{eq:enegw1b}) and ${\mathcal L}$ from (\ref{eq:enegw2}), we obtain
\begin{equation}\label{eq:dl3}
D_L^2\ \frac{1}{2\pi} \frac{c^3}{G} h_\circ \omega_s^2
= \frac{32}{5}\frac{G}{c^5} \mu^2 R^4 \omega_s^6
\end{equation}
(always remembering that $\omega_{gw}=2\omega_s$), and thus
\begin{equation}\label{eq:dl4}
D_L= \frac{8}{\sqrt 5} \frac{G}{c^4} \frac{1}{h_\circ} (\mu R^2 \omega_s^4) \ .
\end{equation}
Let us now insert the value determined in our computation for this event; the reduced mass corresponds to $\mu=17 M_\odot$. 
The values of angular velocity and distance at different $\Delta t $ are reported in Table \ref{tab:chirp}, and the strain $h_\circ$ in Fig. \ref{fig:chirp}. 
We insert into Eq. (\ref{eq:dl4}) the values corresponding to $\Delta t=16.0$ ms: 
$h_\circ\simeq 0.8\times 10^{-21}$,
$\omega_s = \pi \nu_{gw} = \pi\ 63$ Hz  $\simeq 200$ s$^{-1}$,
$R=4.7\times 10^5$ m. We obtain
\begin{equation}\label{eq:dl5}
D_L \simeq 1.1\times 10^{25} \textrm{ m} = 0.4  \textrm{ Gpc} \ ,
\end{equation}
value in agreement with the luminosity distance of 410 Mpc  reported in the paper (notice the large error on this estimate).

The redshift of an object cannot be directly measured using GWs. If the source producing the GW is identified through a different measurement (as part of a multimessenger program, as we will see for the case discussed in Sect.  \ref{sec:gw1708}), the redshift measured with different instruments can be used. Otherwise (as in the case of GW150914), the $z$ can be determined assuming standard cosmology (see for instance \S 21. Big-Bang cosmology of \cite{pdg}, Fig. 21.1). For a luminosity distance of $D_L\simeq 400$ Mpc, the corresponding redshift is $z\simeq 0.1$.
From such (relatively) small redshift value, the relation (\ref{eq:dl3}) is affected by a correction smaller than the uncertainties on the measured quantities. 
The quantity that can be measured with a relatively small uncertainty is the chirp mass, and this value can be corrected for the redshift, as shown below. 

Like the electromagnetic radiation, GWs are stretched by the expansion of the Universe. This increases the wavelength (at \textit{redshift} z), decreases the frequency of the waves detected (``det'') on Earth compared to their values when emitted at the source (``s'') and time intervals are ``redshifted'' at the location of the observer as 
\begin{subequations}\label{eq:cosmo1}
\begin{equation}
\Delta t^{det} = (1+z)\Delta t^{s} \ . 
\end{equation}
Thus, redshift has the following effects on observables:
\begin{equation}
\nu^{det} = \frac{\nu^s}{1+z} \ ,
\end{equation}
\begin{equation}
{\dot \nu^{det}}= \frac{\Delta\nu^{det}}{\Delta t^{det}}=
\frac{\Delta\nu^{s}}{\Delta t^{s}} \frac{1}{(1+z)^2} =
{\dot \nu^{s}}  \frac{1}{(1+z)^2} \ .
\end{equation}
\end{subequations}
The effect on the chirp mass \textit{at the source frame} can be derived using Eq. (\ref{eq:ene10}), which correspond to the \textit{detected} value:
\begin{align}\label{eq:cosmo2}
{\mathcal M}^{det} & \propto (\nu_{gw}^{det})^{-11/3}\ (\dot \nu_{gw}^{det})^{3/5} \nonumber \\ 
&= \frac{(\nu^s)^{-11/5}}{(1+z)^{-11/5}}
\frac{(\dot \nu^s)^{3/5}}{(1+z)^{6/5}} 
= (1+z) (\nu^s)^{-11/5} (\dot \nu^s)^{3/5} \nonumber \\
{\mathcal M}^{det} & = (1+z) {\mathcal M}^{s} \ .
\end{align}
Consequently, the individual masses of the involved objects as measured on Earth are scaled up by a similar factor as the chirp mass:
\begin{equation}\label{eq:cosmo3}
m_1^{det} = (1+z) m_1 \quad ; \quad m_2^{det} = (1+z) m_2 \ ,
\end{equation}
as can be easily verified from the definition of chirp mass, Eq. (\ref{eq:ene8}). The direct inspection of the detector data yields mass values from the red-shifted waves, and thus the values we derived in (\ref{eq:co5}) must be scaled down by $(1+z)$ to obtain the values at the source frame (those reported in the abstract of the paper).

In conclusion, from the derived redshift of $z\simeq 0.1$, the masses at the source frame are about 10\% smaller than that derived in (\ref{eq:co5}) at the detector frame.

\subsection{Total emitted energy\label{sec:eloss}}

Another impressive observation of the binary BH merger is the surprising amount of energy emitted in the form of gravitational radiation by GW150914.

We can evaluate the total gravitational energy radiated starting from the value of the total energy of the orbiting BHs given by (\ref{eq:ene1}).
We assume an initial very large distance of the black holes, $R \rightarrow \infty$, and a final separation given by the sum of their Schwarzschild radii, Eq. (\ref{eq:co1}). From this, we have
\begin{equation}\label{eq:eloss1}
\Delta E=E_{tot}^f- E_{tot}^i= -\frac{G m_1 m_2}{{2\mathcal R}} =
-\frac{Gm_1 m_2 c^2}{{4GM}} 
= \frac{\mu c^2}{4} \simeq 4 M_\odot c^2
\end{equation}
or $5\times 10^{47}$ J, as the estimate of the total amount of gravitational wave energy radiated, in agreement with the value of $3M_\odot$ c$^2$ determined in \cite{1}. 
Equation (\ref{eq:eloss1}) also shows that for a fixed total mass $M= m_1+ m_2$, the radiated energy depends on the reduced mass $\mu$ of the system, and thus it is maximum when the merging BH masses are equal.

This enormous amount of energy is emitted, according to the waveform of Fig. \ref{fig:chirp}, in a tenth of a second.
During the $10^{10}$ y of lifetime, a star like the Sun is expected to convert less than 1\% of its mass into light and radiation.
Thus, the energy emitted by the two BHs during $\sim$ 0.1 s as GWs is $\sim 300$ times as much energy as the electromagnetic radiation emitted by the Sun during its history.

\subsection{Ringdown stage: Spin of the BHs\label{sec:spin}}

The above Newtonian approximations ignore polarization of the gravitational radiation and the intrinsic angular momentum (spin) of the BHs.
Their spin leads to additional velocity-dependent interactions during inspiral. This is analogous to that acting on satellites and gyroscopes in the Earth orbit, due to the rotation of the Earth. 
For binary systems (BHs or NSs) undergoing inspiral these forces are much more important due to the larger masses and (almost) relativistic velocities involved.
Incorporation of these effects and other refinements is not straightforward in terms of an elementary presentation.
 
For an object with mass $m$ and spin $\vec{S}$, the dimensionless spin parameter is defined as
\begin{equation}\label{eq:chi}
 \chi = \frac{c}{G} \frac{|\vec{S}|}{m^2} \ .
\end{equation}
The spin modify the radius of the event horizon with respect to the Schwarz\-schild radius: for an object with $\chi=1$, the event horizon correspond to $Gm/c^2$, half of the value of $\mathcal{R}$ for a non-spinning BH.
Thus, for two $\chi>0$ rotating BHs the system is more compact than for $\chi=0$ objects. The spins of the initial BHs can be inferred using templates modelled on the inspiral data. From this, the LIGO/Virgo Collaboration determined that the spin of the primary BH (the more massive) is constrained to have $\chi < 0.7$, while the spin of the secondary is only weakly constrained.

The effects introduced by the BH spin is more important in the third and final stage, called {ringdown}.
During this stage, the merged object relaxes into its equilibrium state, a Kerr black hole.
The ringdown process can still be analytically treated with general relativity formulas.
As mentioned, during the ringdown phase, the strain $h$ in Fig. \ref{fig:chirp} looks like the transients of a damped harmonic oscillator (the ``quasinormal'' mode).
The damping rate and ringing frequency of the quasinormal mode depend only on the mass and spin of the quiescent Kerr BH that forms after the merging.

The final spin of the black hole was estimated with $\chi=0.67^{+0.05}_{-0.07}$.
Thus, the spins of the initial BHs, determined using the inspiral data, and the spin of the final merged object, determined using a numerical analysis of the ringdown, agree each other. 
Although still with large uncertainty, this result represents the first experimental test of general relativity in the hitherto inaccessible \textit{strong field regime}, and it constitutes another significant outcome of the LIGO/Virgo discovery.

\subsection{Source Localization in the Sky \label{sec:loca}}
Gravitational wave interferometers are linearly-polarized quadrupolar detectors and do not have good directional sensitivity. 
As a result, two antennas are necessary in order to obtain minimum directional information on the source position using the relative arrival time of the signal. 
The two LIGO antennas have a separation baseline of $L \sim 3 \times 10^6$  m; thus, the gravitational wave at 200 Hz (the frequency at which the signal has maximum strain) has wavelength $\lambda =1.5\times 10^6$ m, and thus the detector has a resolution of 
\begin{equation}\label{eq:res1}
\Delta \theta \simeq \frac{\lambda}{L} = 0.5 \textrm{ rad}\ \sim 28^\circ \ .
\end{equation}
The uncertainty on the source position corresponds to about $\Delta \theta^2 \sim 800$ deg$^2$. The 90\% credible region mentioned in \cite{1} corresponds to approximately 600 deg$^2$. 
The localization improves significantly using three detectors.
By measuring the time differences in signal arrival times at various detectors in a network (triangulation technique), the $\Delta \theta^2$ reduces by an order of magnitude or more.

\section{GW170817, GRB170817 and AT 2017gfo}
\label{sec:gw1708}

If sufficiently close to the Earth, the merger of two neutron stars (NSs) is predicted to produce three observable phenomena: a GW signal; a short burst of $\gamma$-rays (GRB) and, possibly, neutrinos; a transient optical-near-infrared source.
Such transient (called also ``kilonova'') would be powered by the synthesis of large amounts of very heavy elements such as gold and platinum via rapid neutron capture (the so-called astrophysical \textit{r-process}.)

\begin{figure}[tb]
\begin{center}
\includegraphics[width=11.0cm]{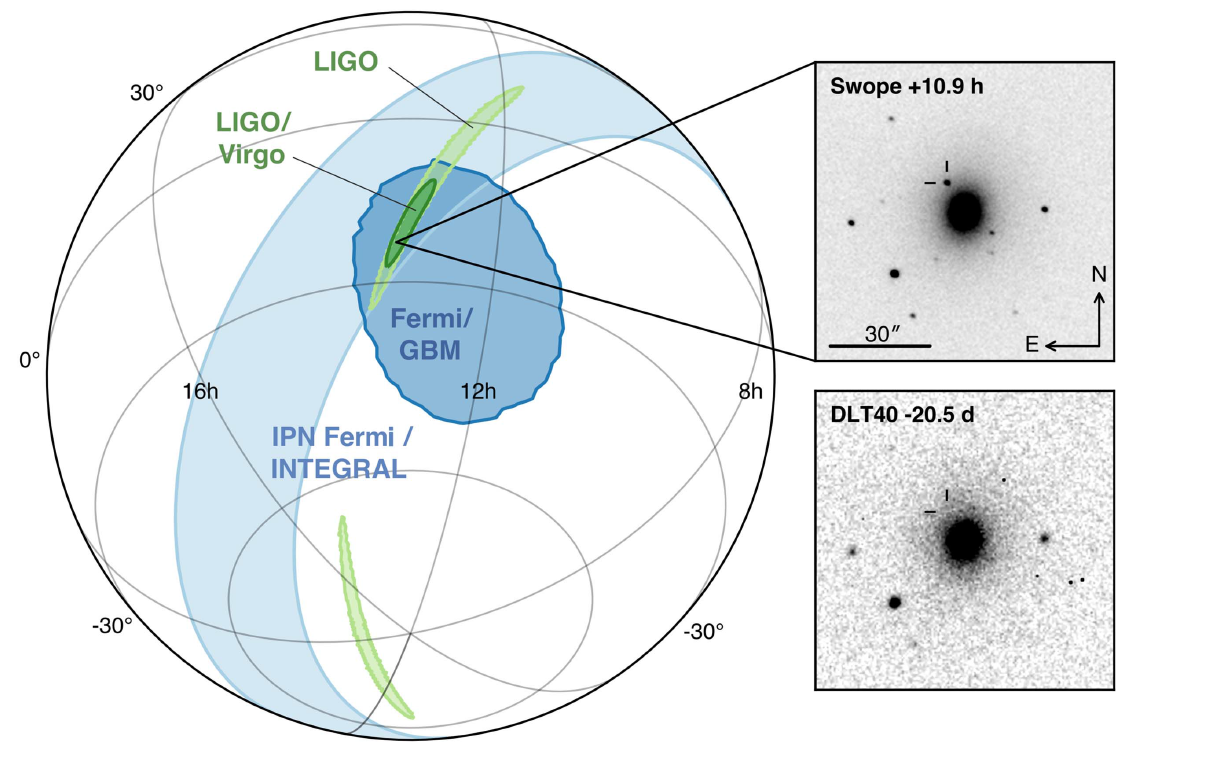}
\end{center}
\caption{\small\label{fig:loca} 
Localization of the GW, GRB, and optical signals. On the left, the orthographic projection of the 90\% credible regions from LIGO alone (190 deg$^2$, light green); the initial LIGO-Virgo localization (31 deg$^2$, dark green); the result from the triangulation from the time delay between Fermi and INTEGRAL (light blue); and Fermi-GBM (dark blue). The inset shows the location of the apparent host galaxy NGC 4993 in the Swope optical discovery image 11 hours after the merger (top right). Below, the pre-discovery image from 20.5 days prior to merger from another telescope, the DLT40 (bottom right). The reticle marks the position of the transient in both images. From \cite{3}.}
\end{figure}

On August 17, 2017, 12:41:04 universal time (UT) the LIGO-Virgo detector network observed a GW signal from the inspiral of two low-mass compact objects consistent with a binary NS merger (GW170817).
Independently, a $\gamma$-ray burst (GRB170817A) was observed less than 2 s later by the Gamma-ray Burst Monitor on board the Fermi satellite, and by INTEGRAL satellite.
This joint GW/GRB detection was followed by the most extensive worldwide observational campaign never performed before, with the use of space- and ground-based telescopes, to scan the sky region where the event was detected. The localization on the sky of the GW, GRB, and optical signals is presented in Fig. \ref{fig:loca}. 
Also underwater/ice neutrino telescopes looked for a neutrino counterpart of the signal.  
Less than 12 h later (without the Sun on the signal region), a new point-like optical source was reported by different optical telescopes. The source was located in the galaxy NGC 4993 at a distance of 40 Mpc from Earth, consistent with the luminosity distance of the GW signal. Its official designation in the International Astronomical Union (IAU) is AT 2017gfo. The source was intensively studied in the following weeks by all traditional astronomical instruments from radio to X-rays. 
The interest and effort have been global: a large number of papers on different observations was published on the same issue of 
The Astrophysical Journal Letters (Vol. 848, n. 2) on October 20, 2017.
This includes one paper describing the multi-messenger observations \cite{3} which is coauthored by almost 4,000 physicists from more than 900 institutions, using 70 observatories on all continents and in space, see Fig. \ref{fig:1gw}.

\subsection{GW170817}
\label{sec:gw17}

Binary NS systems produce GWs with luminosity (in the Newtonian approach) given by Eq. (\ref{eq:enegw2}). 
As the orbit of a binary NS system gets smaller, the GW luminosity increases, accelerating the inspiral. This process has long been predicted to produce a GW signal observable by ground-based detectors in the final minutes before the massive objects collide.
To give an idea of astrophysical uncertainties, models of the population of compact binaries predicted for the network of advanced GW detectors a number of possible observations ranging from ${\mathcal O}(0.1)$ to ${\mathcal O}(100)$ every year.  

The first indirect observation of a binary NS system releasing energy in form of gravitational radiation comes in 1974 with the discovery of the first system with two rotating NSs by Hulse and Taylor. They found that this binary NS system was losing energy at a rate equal to that foreseen by the emission of gravitational waves.

\vskip 0.3cm
\noindent\textbf{Exercise: The Hulse and Taylor pulsar.} 
{\small PSR B1913+16 is a pulsar which together with another NS is in orbit around a common center of mass, thus forming a binary star system. It is also known as \textit{Hulse–Taylor binary system} after its discoverers.
The period of the orbital motion is $T=7.7517$ hours, and the period decay with a rate of $\dot T=(-3.2\pm 0.6)\times 10^{-12}$ s s$^{-1}$.

\noindent 1) Compute the energy emitted by the system, assuming $m_1=m_2=1.4 M_\odot$ and a circular orbit. 2) Estimate the decay rate of the period, $\dot T $, assuming emission of GWs. 

\noindent The above estimate needs to be revised to allow the non-negligible eccentricity of the orbit, $\epsilon = 0.617$. This yields an additional multiplicative factor on $\mathcal L$ given by $f(\epsilon)= (1+7/24 \epsilon^2 + 37/96 \epsilon^4)(1-\epsilon^2)^{-7/2}$ (see \cite{schutz}). 
The factor $f(\epsilon=0.617)= 12$  explains why the orbit of binary systems are circular before merging. 
The luminosity $\mathcal L$ depends on the angular velocity of the system to a high power, and the system rearrange its orbit to a circular one to minimize the energy loss in term of gravitational radiation. }

\vskip 0.3cm

Toward the end of the data run O2 of aLIGO and aVirgo, a binary NS signal, GW170817, was identified by matched filtering the data against post-Newtonian waveform models. 
The signal was observed for about $\sim $ 100 s in the sensitive frequency band of GW interferometers (at frequency $>$ 24 Hz) then, the inspiral signal ended at 12:41:04.4 UTC. 
During the few minutes needed by the matched filters to pick-up the signal from data stream, a $\gamma$-ray burst (GRB) was observed and reported by satellites. The GRB occurred 1.7 s after the coalescence time, derived by the GW signal. 
The combination of data from GW detectors allowed a sky position localization to an area of 28 deg$^2$ within few hours, enabling the electromagnetic follow-up campaign that identified a optical counterpart in the galaxy NGC 4993. 
\begin{figure}[tb]
\begin{center}
\includegraphics[width=12.0cm]{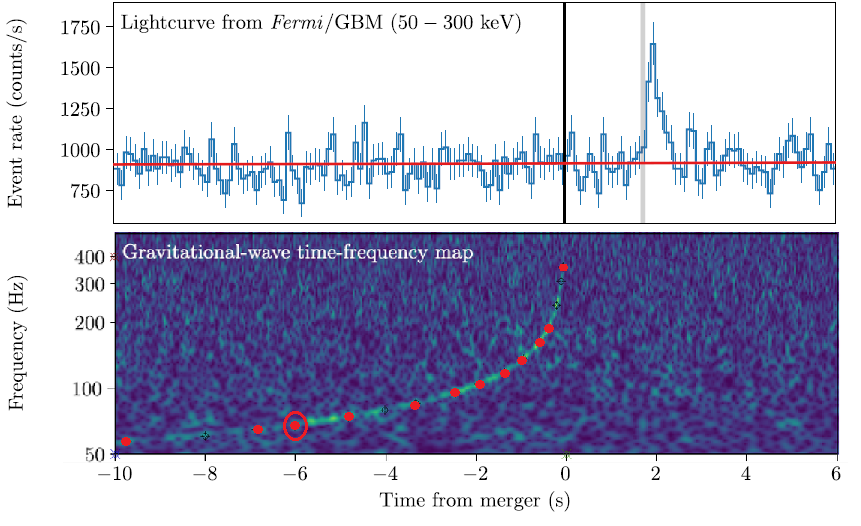}
\end{center}
\caption{\small\label{fig:chirp2} 
Part of the joint, multimessenger detection of GW170817 and GRB170817A. Top: the summed Fermi-GBM light curve in the 50-300 keV energy range. Bottom: the time-frequency map of GW170817. 
All times here are referenced to the GW170817 coalescence time $T_0$. The markers on the bottom panel (one of which is highlighted by a red circle) are used in the present analysis to infer the frequency-time values reported in Table \ref{tab:chirp2}. Adapted from \cite{4}.}
\end{figure}

The time evolution of the frequency of the GW emitted by a binary NS system before merging is determined primarily by the chirp mass, Eq. (\ref{eq:ene8}). 
We can estimate $\mathcal M$, according to Eq. (\ref{eq:ene9}), extracting numerical values from the time-frequency representation of the signal shown in the bottom panel of Fig. \ref{fig:chirp2}. 
Tab. \ref{tab:chirp2} reports, for different $\Delta t$ from time of the coalescence, the derived values of chirp mass $\mathcal M$ and radius $R$ of the system. 

\begin{table}[t]
\caption{\small Distance $\Delta t$ from time of the coalescence and frequency $\nu_{gw}$ as obtained from GW time-frequency map of Fig. \ref{fig:chirp2}; from the third column: the frequency change rate, $\dot \nu_{gw}$, evaluated as $\Delta \nu_{gw}/\Delta t_{gw}$; the value of the chirp mass, ${\mathcal M}$, as obtained from Eq. (\ref{eq:ene10}); the distance $R$ between the two NSs evaluated with the Kepler's third law (remember: $\omega_s = \pi \nu_{gw}$) }\label{tab:chirp2}
\begin{center}
\begin{tabular}{cccccc}
\hline
$\Delta t$ & $\nu_{gw}$ & $\dot \nu_{gw}$ & ${\mathcal M}$ & ${\mathcal M}/M_\odot$ & R \\
(s)       & (Hz) & (Hz s$^{-1})$ & (kg) & & (km) \\
\hline
-9.74& 57.1	 & -   & -  & - & 	166 \\
-6.87	&64.8	&2.7 &	2.1$\times 10^{30}$	&1.0	&153\\
-4.83	&74.3	&4.7	&2.2$\times 10^{30}$	&1.1	&140\\
-3.33	&85.7	&7.6	&2.1$\times 10^{30}$	&1.1	&127\\
-2.45	&95.7	&11.4	&2.1$\times 10^{30}$	&1.1	&118\\
-1.93	&104.7 &17.2	&2.2$\times 10^{30}$&	1.1	&111\\
-1.37	&118.2 &23.8	&2.1$\times 10^{30}$	&1.0	&102\\
-0.94	&136.3 &42.8	&2.1$\times 10^{30}$	&1.1	&93\\
-0.59	&163.1 &75.1	&2.0$\times 10^{30}$	&1.0	&83\\
-0.21	&239.7 &201.1	&1.6$\times 10^{30}$	&0.8	&64\\
-0.06	&359.9 &810.0	&1.5$\times 10^{30}$	&0.7	&49\\
\hline
\end{tabular}
\end{center}
\end{table} 

As the orbital separation $R$ approaches the size of the bodies, the gravitational wave is increasingly influenced by relativistic effects related to the mass ratio $q = m_2/m_1$, where $m_1 \ge m_2$, as well as spin-orbit and spin-spin couplings.
You can notice that in the last rows of Tab. \ref{tab:chirp2} the derived value of the chirp mass differs from the values at early times.
This means that the details of the objects' internal structure become important. 
For neutron stars, the tidal field of the companion induces a mass-quadrupole moment and accelerates the coalescence.
As a BH has no-hairs, tidal effects have not been considered in the above discussion of GW150914. 
The \textit{tidal polarizability parameters} are important because contain information on the nuclear equation of state for NSs (see below).

As for GW150914, the properties of GW sources have been inferred by matching the data with predicted waveforms. The results of the LIGO/Virgo collaboration, reported in Table \ref{tab:gw170817} and discussed below, include dynamical effects from tidal interactions, point-mass spin-spin interactions, and couplings between the orbital angular momentum and the orbit-aligned dimensionless spin components of the stars, $\chi$. 

\begin{table}[tb]
\caption{\small Source properties for GW170817. The central values encompass the 90\% credible intervals for different assumptions of the waveform model to bound systematic uncertainty. The masses are quoted in the frame of the source, accounting for uncertainty in the source redshift. Adapted from \cite{3}} \label{tab:gw170817}
\begin{center}
\begin{tabular}{lcc}
\hline
& $|\chi_{NS}|<0.05$ & $|\chi_{NS}|<0.89$ \\
\hline
Chirp mass $\mathcal M$  & $1.188^{+0.004}_{-0.002}\ M_\odot$ & $1.188^{+0.004}_{-0.002}\ M_\odot$ \\
Luminosity distance $D_L$ & $40^{+8}_{-14}$ Mpc &$40^{+8}_{-14}$ Mpc \\
Mass ratio $q=m_2/m_1$& 0.7-1.0 & 0.4-1.0 \\
Total mass $M=m_1+m_2$& $2.74^{+0.04}_{-0.01} M_\odot$ &$2.82^{+0.47}_{-0.09} M_\odot$ \\
Primary mass $m_1$   & 1.36-1.60 $M_\odot$ & 1.36-2.26 $M_\odot$ \\
Secondary mass $m_2$ & 1.17-1.36 $M_\odot$ & 0.86-1.36 $M_\odot$ \\
Viewing angle $\Theta$ & $ \le 55^\circ$ & $ \le 56^\circ$ \\
Using NGC 4993 location & $ \le 28^\circ$ & $ \le 28 ^\circ$ \\
Tidal deformability $\Lambda(1.4 M_{odot})$ & $\le 800$ & $\le 1400$\\
Radiated energy $E_{rad}$ & $> 0.025\ M_\odot c^2$ & $> 0.025\ M_\odot c^2$ \\
\hline
\end{tabular}
\end{center}
\end{table} 

\vskip 0.2cm
\noindent\textbf{Chirp mass.} 
Our simple Newtonian approach gives in Table \ref{tab:chirp2} a value of ${\mathcal M} \sim 1.1 M_\odot$ (a part the last two rows).  
In the detailed analysis of \cite{2}, the chirp mass is the best-determined quantity. The value obtained from the GW phase, ${\mathcal M}^{det}= 1.1977 M_\odot$ correspond to the detector frame, and it is related to value assumed at the rest-frame of the source by its redshift $z$ as given in (\ref{eq:cosmo2}).
A redshift of $z=0.008$ is derived from the luminosity distance and the cosmology parameters, which is consistent with the known distance of galaxy NGC 4993.
The values of masses reported in Table \ref{tab:gw170817} are corrected for this redshift value.

\vskip 0.2cm
\noindent\textbf{Luminosity distance.} According to the discussion in \S \ref{sec:cosmo}, the luminosity distance $D_L$ can be obtained from the masses of the system and the strain $h$. In the case of GW170817, $h\sim 10^{-22}$ and $D_L$ is obtained with a 20\%-30\% uncertainty. Refer to Eq. (\ref{eq:bygw11}) which uses the values derived from this NS system.

\vskip 0.2cm
\noindent\textbf{Individual masses: mass ratio and total mass.}
While $\mathcal M$ is well constrained, the estimates of the component masses are affected by the degeneracy between mass ratio $q$ and the aligned spin components of the two NSs.
These latter values are very poorly constrained from data, also combined with external information about the total angular momentum, $\textbf{J}$,  of the system. 
In fact $\textbf{J}$ corresponds to the sum of the orbital angular momentum of the two rotating masses and the individual spins of the NSs. 
Due to low masses of NSs, the NS spins have little impact on the total angular momentum.
While the dimensionless spin parameter (\ref{eq:chi}) assumes values $\chi<1$ for black holes, realistic NS equations of state typically imply $\chi<0.7$. Thus, in Table \ref{tab:gw170817}, two different assumptions (or ``priors'') have been considered: a high-spin value ($|\chi_{NS}|\le 0.89$) and a low-spin value ($|\chi_{NS}|\le 0.05$).
The mass ratio, $q=m_2/m_1$, changes according to these two priors.
The central values of the total mass, $M$, of the system are very close in the two cases and always compatible with the presence of two equal objects with masses close to $1.4 M_\odot$. 

\vskip 0.2cm
\noindent\textbf{Inclination angle.} 
The total angular momentum, $\textbf{J}$, is (almost) perpendicular to the plane of the orbit. 
The luminosity distance is correlated with the inclination angle 
\begin{equation}\label{eq:ns3}
\cos\theta_{JN} = \frac{{\bf J}\cdot \hat{\bf N}}{J} \ ,
\end{equation}
where $\hat{\bf N}$ is the unit vector from the source towards the
Earth. 
Data are consistent with an antialigned source: $\cos\theta_{JN}\le -0.54^\circ$. The relevant quantity is the viewing angle 
\begin{equation}\label{eq:ns4}
\Theta \equiv min(\theta_{JN}; 180^\circ - \theta_{JN}) \ ,
\end{equation}
which corresponds, in this case, to $\Theta \le 56^\circ $. However, since $D_L$ can be determined using the multimessenger association with the galaxy NGC 4993, Eq. (\ref{eq:ns3}) can be further constrained to $\cos\theta_{JN}\le -0.88^\circ$ and thus $\Theta \le 28^\circ $.

\vskip 0.2cm
\noindent\textbf{Tidal deformability and energy emitted in GW.} 
Tides are well known effects in the study of planet's motions. As early as in the 1910s, Augustus E. Love introduced two dimensionless parameters ($k_1, k_2$) to characterize the rigidity of a planetary body and the susceptibility of its shape to change in response to a tidal potential.
In particular, $k_2$ encodes information about the body's internal structure and it is defined as the ratio between the tidally induced quadrupole moment $Q_{ij}$ and the companion's perturbing tidal gradient (the external field). 
The tidal deformability (or polarizability) is:
\begin{equation}\label{eq:ns1}
\Lambda = \frac{2}{3} k_2 \biggl(\frac{c^2}{G}\frac{R}{m}\biggr)^5 
\end{equation}
(we do not give any derivation of this; see \cite{2} and referred papers).
Both $R$ (the stellar radius) and $k_2$ are fixed for a given stellar mass $m$ by the equation of state (EOS).
For neutron-star matter (according to the discussion in \cite{2}) $k_2 \simeq 0.05-0.15$ while black holes have $k_2=0$. 
Tidal effects increasingly affect the phase of the GW and become significant above $\nu_{gw}\simeq 600$ Hz, so they are potentially observable in ground-based interferometers. 
Unfortunately, interferometers in the O2 run were not sufficiently sensible above 400 Hz.  

Gravitational wave observations alone are able to set a lower limit on the compactness of the NS system and provide information on the equation of state (EOS) through an estimate of the deformability (\ref{eq:ns1}).
The values of $\Lambda$ for GW170817 reported in the table disfavor EOS predicting less compact stars; objects more compact than neutron stars such as quark stars, black holes, or more exotic objects are not excluded. 
The energy emitted, $E_{rad}$, depends critically on the EOS.
For this reason, only a lower bound on the energy emitted before the onset of strong tidal effects at $\nu_{gw}\sim 600$ Hz is derived, which is consistent with that obtained from numerical simulations.

\begin{figure}[tb]
\begin{center}
\includegraphics[width=11.0cm]{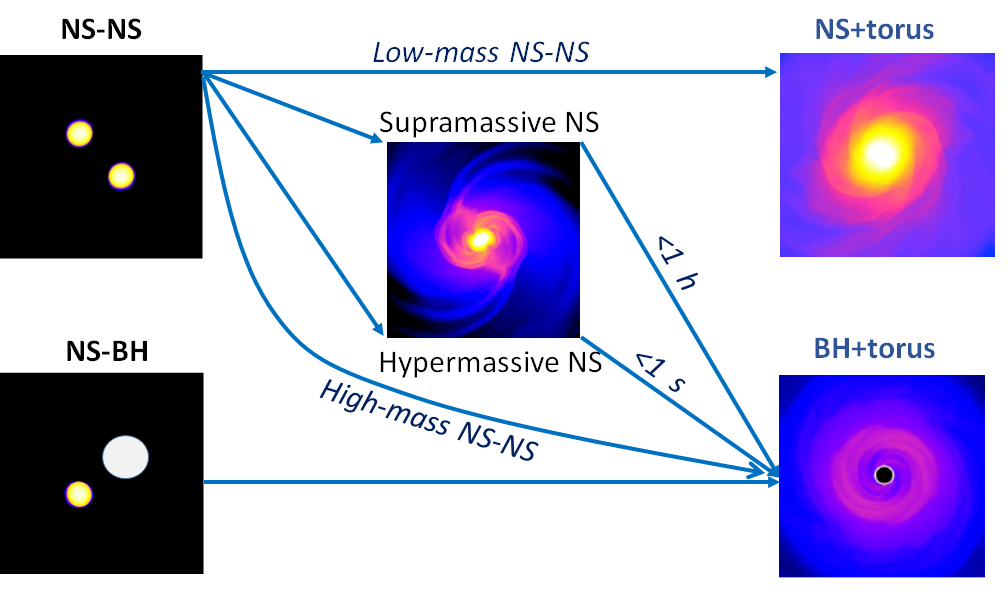}
\end{center}
\caption{\small\label{fig:final} 
Possible final state of a binary NS-NS or NS-BH system.}
\end{figure}

\vskip 0.2cm
\noindent\textbf{Final state after the collision.}
One interesting subject (not presented in the discovery paper and in Table \ref{tab:gw170817}) is the fate of the system after the collision \cite{final}.
After such a merger, a compact remnant is left over whose nature depends primarily on the masses of the inspiralling objects and on the EOS of nuclear matter. 
This could be either a BH or a NS, with the latter being either long-lived or too massive for stability implying delayed collapse to a BH (Fig. \ref{fig:final}). Depending on the mass of the intermediate state (hypermassive NS or supramassive NS), short ($<1$ s) or intermediate-duration ($<500$ s) GW emission is expected. No signal was found in this case, so no particular mechanism for the formation of the final state is defined. 
However, models shows that post-merger emission from a similar event may be detectable when advanced detectors reach design sensitivity or with next-generation detectors.

\subsection{GRB170817A}
\label{sec:grb17}
Gamma-ray bursts (GRBs) are extremely intense and relatively short bursts of gamma radiation observed by dedicated satellite experiments, coordinated in the Gamma-ray Coordinates Network (GCN)\footnote{\url{https://gcn.gsfc.nasa.gov/}}. 
The GCN system provides to distribute the locations of GRBs and other transients detected by spacecraft. Most alerts are in real-time while the burst is still bursting and others are delayed due to telemetry down-link delays. GRBs are reported at a rate of one or two per day. 
The GCN reports also of follow-up observations (the Circulars) made by ground-based and space-based optical, radio, X-ray, TeV $\gamma$-rays, and other particle observers \footnote{The GCN circulars for GW170817 follow-up are available in \url{https://gcn.gsfc.nasa.gov/other/G288732.gcn3}}. 

In a GRB, after the initial flash of $\gamma$-rays, a longer-lived {}``afterglow'' is usually emitted at longer wavelengths (X-ray, ultraviolet, optical, infrared, microwave and radio).
Since the observation of first afterglow from the Beppo-SAX satellite in 1997, we know that GRBs are of extragalactic origin and that they are the brightest electromagnetic events known to occur in the Universe.

GRBs are classified as $short$ ($\Delta t\le 2$ s) or $long$ ($\Delta t>2$ s) depending on the duration $\Delta t$ of their prompt $\gamma$-ray emission. This division is based on the observed bimodal distribution of $\Delta t$ and on differences in the $\gamma$-ray spectra. 
This empirical division was accompanied by hypotheses that the two classes have different progenitors. 
Long GRBs have been firmly connected to the collapse of massive stars through the detection of associated Type Ibc core-collapse supernovae. Prior to GRB170817A, the connection between short GRBs and mergers of binary NSs (or NS-BH binaries) have been supposed by numerical simulations and have only weak indirect observational evidence.

GRBs are thought to be produced by internal shocks in relativistic expanding \textit{fireballs}. A fireball, the relativistic outflows or jets of plasma, is created when the central engine releases a large amount of energy over a short time and small volume. This is the case of the merging of a binary NS system. 
These jets are launched along the rotational axis of the progenitor, powered by the gravitational energy released during temporary mass accretion onto the central black hole.
A GRB (either short or long) consists of a \textit{prompt emission}, followed by several components, such as an \textit{extended emission}, X-ray flares, and plateau emission, which usually are referred for as the \textit{afterglow emission}. The prompt emission is attributed to internal energy dissipation inside the relativistic jet, whereas the afterglows are thought to be caused by forward shocks propagating in the surrounding ambient material. 

GRB170817A was autonomously detected in-orbit by the Fermi-GBM flight software +1.74 s after the GW coalescence (see Fig. \ref{fig:chirp2}). A similar coincidence was observed by an instrument on board the INTEGRAL satellite.
The onset of $\gamma$-ray emission from a binary NS merger progenitor is predicted from models to be within a few seconds after the merger, given 
the expected formation time of the central engine and the jet propagation delays that are of the order of the GRB duration $\Delta t$.
The unambiguous joint detection of GW and electromagnetic radiation from the same event (the occurrence of a observation by chance has estimated probability of $5\times 10^{-8}$) confirms that binary NS mergers are progenitors of (at least some) short GRBs.

The prompt $\gamma$-ray emission from GRB 170817A had an observed energy of $E_{iso} \sim 4\times 10^{46}$ erg, as recorded by Fermi-GBM. The subscript {}``iso'' means that the computation assumes that the gammas are isotropically emitted by the source. 
This is at least three orders of magnitude below typical observed short GRB energies. As discussed below (\S \ref{sec:neutrino}), a plausible explanation is the presence of a beamed emission, with the Earth off-axis with respect to the jet.

\subsection{An (apparent) digression: The origin of the elements}
\label{sec:ele}

\begin{figure}[tb]
\begin{center}
\includegraphics[width=12.0cm]{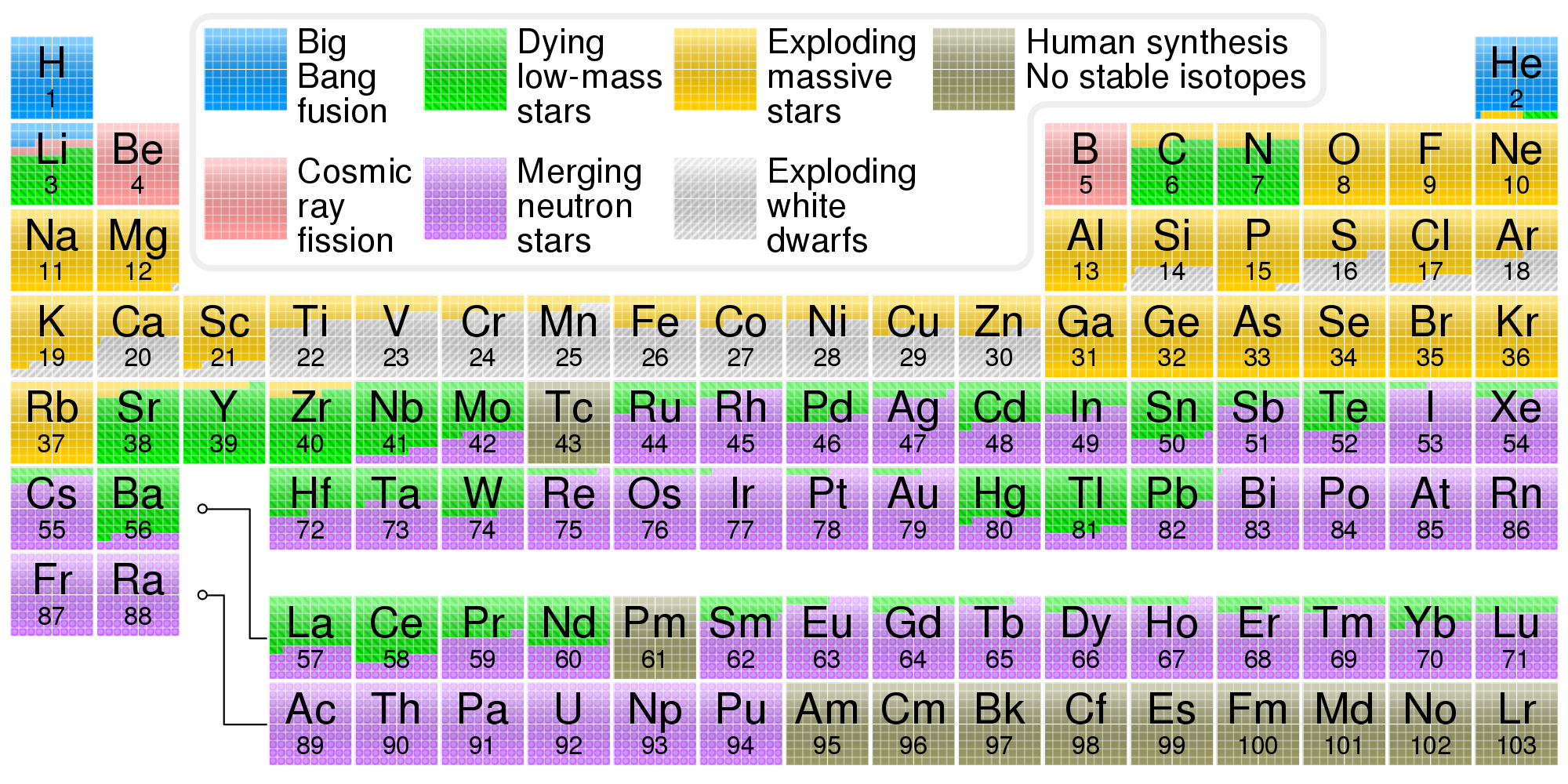}
\end{center}
\caption{\small\label{fig:nucleo} 
A version of the periodic table indicating the main origin of elements found on Earth. The elements with $Z>94$ are mainly of human synthesis.
From \url{https://commons.wikimedia.org/w/index.php?curid=31761437}.}
\end{figure}

One of the most important interconnection between nuclear physics and astrophysics is that needed to explain the origin and the abundance of elements in the Periodic Table (see Fig. \ref{fig:nucleo}).
The abundance of chemical elements in the Universe is dominated by hydrogen and helium, which were produced in the Big Bang. 
Remaining elements, making up only about 2\% of the Universe, have been produced as the results of stellar activities. Nuclear fusion in stars synthetize elements with mass number $A$ up to $56$.
$^{56}$Fe is one of the highest binding energies of all of the isotopes, and is the last element that releases energy by nuclear fusion, exothermically.
Elements of higher mass number become progressively rarer, because they increasingly absorb energy in being produced.
The abundance of elements in the Solar System is thought to be similar to that in the Universe.

The \textit{supernova nucleosynthesis} is the theory of the releasing in the Universe of elements up to iron ($Z=26$) and nickel ($Z=28$) in supernova explosions, first advanced by F. Hoyle in 1954.
Referring to Fig. \ref{fig:nucleo}, the different elements are released in the Universe by different processes. 
Two different exploding stellar scenarios occur. The first involves a \textit{white dwarf} star, which undergoes a nuclear-based explosion after it reaches its Chandrasekhar limit after absorbing matter from a neighboring star. 
The second cause is when a massive star, usually a supergiant, reaches $^{56}$Ni and $^{56}$Fe in its nuclear fusion processes. 

Elements heavier that iron are produced by neutron capture in neutron-rich astrophysical environments, followed by $\beta$ decay, $n\rightarrow p e^- \overline \nu_e$, of some neutrons in the forming nuclei.
The so-called \textit{s-process} is believed to occur mostly in asymptotic giant branch stars. Iron nuclei (the starting material), left by a supernova during a previous generation of stars, capture neutrons produced by the reactions
$^{13}_{\ 6}\textrm{C} +  ^{4}_{2}\textrm{He} \rightarrow ^{16}_{\ 8}\textrm{O} + n$, or 
$^{22}_{10}\textrm{Ne} +  ^{4}_{2}\textrm{He} \rightarrow ^{25}_{12}\textrm{Mn} + n$ occurring in the star. 
The extent to which the s-process moves up the elements in the periodic table to higher mass numbers is essentially determined by the degree to which the star is able to produce neutrons and to the amount of iron in the star's initial abundance. 
The s-process is believed to occur over time scales of thousands of years, passing decades between successive neutron captures.

In contrast, the so-called \textit{r-process} is believed to occur over time scales of seconds in explosive environments. 
The neutron captures must be rapid: the newly formed nucleus does not have time to undergo $\beta$-decay before another neutron arrives to be captured. 
Thus, necessarily the r-process occurs in astrophysical locations where there is a high density of free neutrons. 
Which are those neutron-rich astrophysical regions is a matter of ongoing research. 
Before August 2017, the most suitable candidate was the material ejected from a core-collapse supernova (as part of supernova nucleosynthesis).
GW170817 showed that probably the most suited ambient for r-processes is the neutron-rich matter thrown off from a binary neutron star merger (the \textit{kilonova}).

\subsection{The Kilonova: electromagnetic follow-up of AT
2017gfo}
\label{sec:at17}

\begin{figure}[tbh]
\floatbox[{\capbeside\thisfloatsetup{capbesideposition={right,top},capbesidewidth=6.0cm}}]{figure}[\FBwidth]
{\caption{\small Composition of spectra from the near ultraviolet to the near infrared taken using the X-shooter instrument on ESO's Very Large Telescope. It shows the changing behavior of the kilonova AT 2017gfo over a period of $\sim$11 days after the NS merging. The ejecta is optically thick early on, with a speed of $\sim 0.2c$. 
As the ejecta expands, broad absorption-like lines appear on the spectral continuum, indicating atomic species produced by nucleosynthesis.
A fraction of the synthesized atoms is radioactive; while decaying they heat the ejecta, which then radiates thermally. All the atomic species present in the ejecta have various degrees of excitation and ionization; the absorption from the continuum cause the formation of lines. 
The models that aim at reproducing these lines assume a total explosion energy, a density profile and an abundance distribution of the ejecta. 
The spectral characteristics and their time evolution thus result in a good match with the expectations for kilonovae, suggesting that the merger ejected 0.03-0.05 $M_\odot$ of material, including high-opacity lanthanides.
 Refer to \cite{pian} for further details.}
\label{fig:kilo}}
{\includegraphics[width=6cm]{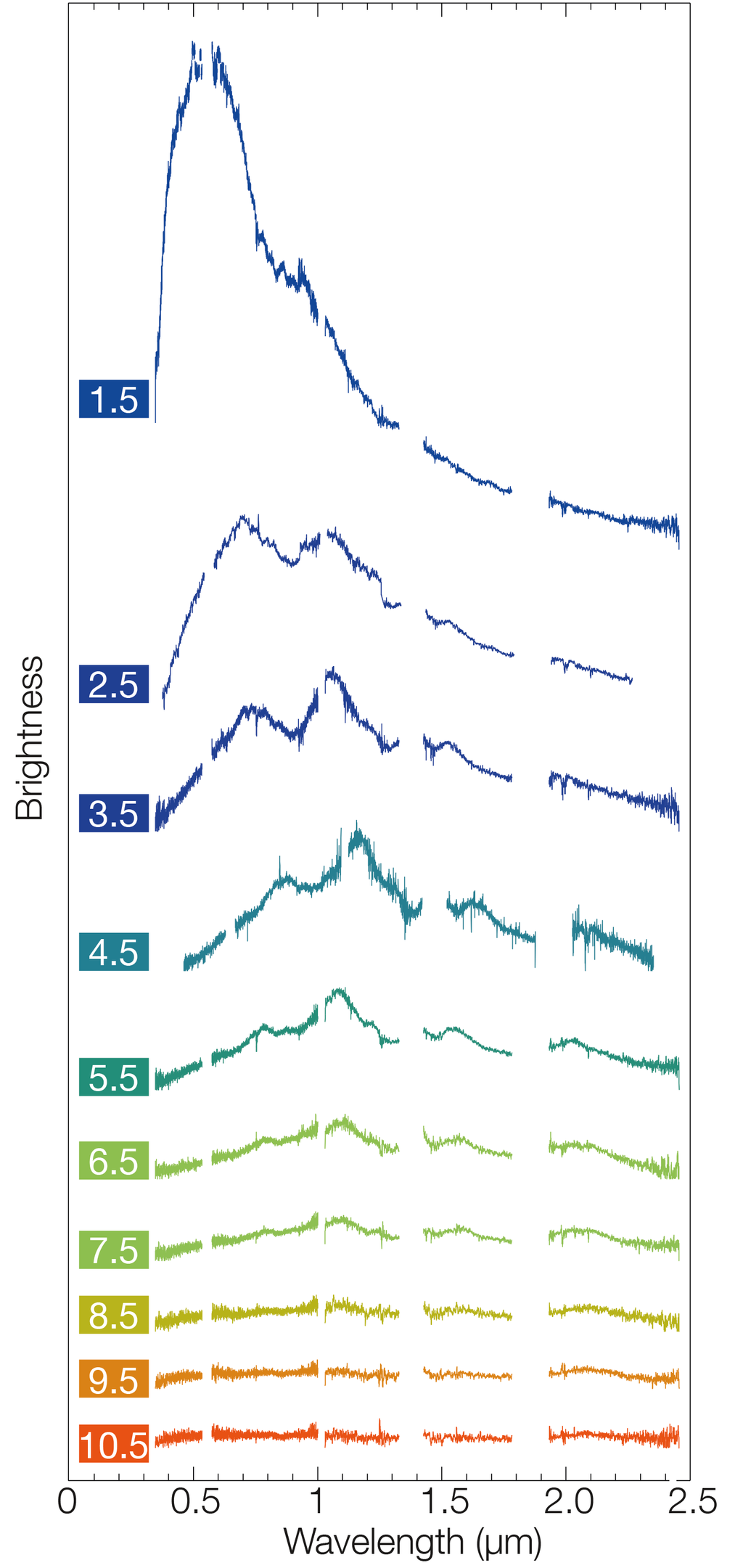}}
\end{figure}

A kilonova is a transient event observable with traditional astronomical methods occurring when two NSs (or a NS-BH system) merge into each other.
The term kilonova (in alternative to \textit{macronova} or \textit{r-process supernova}) was introduced in \cite{Metzger} to characterize the peak brightness of the {isotropic} emission which reaches $10^3$ times that of a classical nova. 
As the authors says in the abstract:
``\textit{Because of the rapid evolution and low luminosity of NS merger transients, electromagnetic counterpart searches triggered by GW detections will require close collaboration between the GW and astronomical communities. ... Because the emission produced by NS merger ejecta is powered by the formation of rare r-process elements, current optical transient surveys can directly constrain the unknown origin of the heaviest elements in the Universe.}'' \cite{Metzger}
This was exactly the situation occurred on August 17$^{th}$, 2017;  the details of the spectral identification and the physical properties of the bright kilonova associated with the GW170817 and GRB170817A are in \cite{pian}.

Following the joint GW/GRB detection, an extensive observing campaign across the electromagnetic spectrum was launched, leading to the discovery less than 11 hours after the merger of a bright optical transient, now with the IAU identification of AT 2017gfo in the galaxy NGC 4993 \cite{3}.
Subsequent observations targeted the object and its environment. Early ultraviolet observations revealed a blue transient that faded within 48 hours. Optical and infrared observations showed an evolution towards red over $\sim$10 days. 

These observations support the hypothesis that, after the merger of two NSs, a kilonova powered by the radioactive decay of r-process nuclei synthesized in the ejecta was produced. 
The information are derived from the series of spectra presented in Fig. \ref{fig:kilo} at different times after the merging from ground-based observatories covering the wavelength range from the ultraviolet to the near infrared.

This multi-wavelength campaign shows that observations are consistent with the presence of an optically thick ejecta at early stages, with a speed of $\sim 0.2c$. 
As the ejecta expands, broad absorption-like lines appear on the spectral continuum, indicating new atomic species synthesized by nucleosynthesis. 
A fraction of the newly formed nuclei is radioactive; their presence is revealed by the fact that, while decaying, they heat the ejecta.
Consequently, the ejected material radiates thermally. 
All the atomic species present in the ejecta have various degrees of excitation and ionization; the absorption from the continuum causes the formation of lines. 
The models that aim to reproduce these lines assume a total explosion energy, a density profile and an abundance distribution of the ejecta. 
The spectral characteristics and their time evolution thus result in a good match with the expectations for kilonovae, suggesting that the merger ejected 0.03-0.05 $M_\odot$ of material, including high-opacity lanthanides \cite{pian}. 

The same conclusion that a minimum of 0.05 $M_\odot$ was produced under the form of heavy elements is independently derived by another analysis \cite{drout}.
Typical solar abundance (by mass fraction) for the r-process elements with mass number A$ > 100$ is $\sim 10^{-7}$.
To explain this value, in our Galaxy, the r-processes need to produce heavy elements at a rate of $\sim 3\times 10^{-7} M_\odot \textrm{ y}^{-1}$ \cite{Metzger}.

If neutron star mergers dominate r-process production over other mechanisms (see Sect. 12.16 of \cite{spurio}), and thus if we assume that all the galactic heavy elements are produced by NS merger events, this production rate requires an event like GW170817 in our Galaxy every 20,000-80,000 years. 
This corresponds to a volume density of such events equal to $(1-4)\times 10^{-7}$ Mpc$^{-3}$ yr$^{-1}$. 
At their design sensitivity, the network of laser interferometers will be able to detect binary NS mergers out to $\sim$ 200 Mpc, leading to a possible detection rate of 3-12 such events per year, or less than one event per year as nearby as GW170817.
If this estimate is correct, in the following few years, we will have an answer about the long-lasting problem of the origin of the heavy elements. 
On the other hand, if the observed rate of GW170817-like events were to be larger, some refinements regarding the theoretical models would be necessary. 
If the GW interferometers end up observing fewer events, other r-process mechanisms will probably have to be considered.  

\section{Astrophysics of stellar BHs after GW150914
\label{sec:astrobh} }

\begin{table}[tbh]
\caption{ Summary table of binary BHs merger detected in O1 (the first three) and O2 runs by the LIGO/Virgo Collaboration. $m_1$ and $m_2$ are the initial masses, ${\mathcal M} $ the chirp mass and $M_{fin}$ the BH mass after merging (all in units of solar mass, $M_\odot$); $\Delta E$ is the total emitted energy as GWs. The parameter $\chi$ is dimensionless spin of the final state BH, $D_L$ is the luminosity distance (in Mpc),  and $\Delta \Omega$ the sky localization. Notice the improvement in this latter parameter when Virgo joined the LIGO operations early in August 2017. From  \cite{LVBH1}.}\label{tab:summ}
\begin{center}
\begin{tabular}{c|cccccccc}
\hline
name & $m_1$ & $m_2$ &  ${\mathcal M} $ & $M_{fin}$ & $\Delta E$  &$\chi$  & $D_L$ & $\Delta \Omega$ \\
& $M_\odot$ &$M_\odot$ &$M_\odot$ & $M_\odot$ &  $M_\odot c^2$ & & Mpc & deg$^2$ \\ \hline
GW150914 & 35.6 & 30.6 & 28.6 & 63.1 & 3.1 & 0.69 & 430 & 180 \\
GW151012 & 23.3 & 13.6 & 15.2 & 35.7 & 1.5 & 0.67 &1060 & 1555\\
GW151226 & 13.7 & 7.7  & 8.9  & 20.5 & 1.0 & 0.74 & 440 & 1033 \\
GW170104 & 31.0 & 20.1  & 21.5  & 49.1 & 2.2 & 0.66 & 960  & 924\\
GW170608 & 10.9 & 7.6 & 7.9 & 17.8 & 0.9 & 0.69 & 320 & 396\\
GW170729 & 50.6 & 34.3& 35.7 & 80.3 & 4.8 & 0.81 & 2750 & 1033\\
GW170809 & 35.2 & 23.8& 25.0 & 56.4 & 2.7 & 0.70 & 990 & 340\\
GW170814 & 30.7 & 25.3 & 24.2 & 53.4 & 2.7 & 0.72 & 580  &  87 \\
GW170818 & 35.5 & 26.8 & 26.7 & 59.8 & 2.7 & 0.67& 1020 & 39\\
GW170823 & 39.6 & 29.4 & 29.3 & 65.6 & 3.3 & 0.71& 1850 & 1651\\
\hline
\end{tabular}
\end{center}
\end{table} 

\begin{figure}[tb]
\begin{center}
\includegraphics[width=11.7cm]{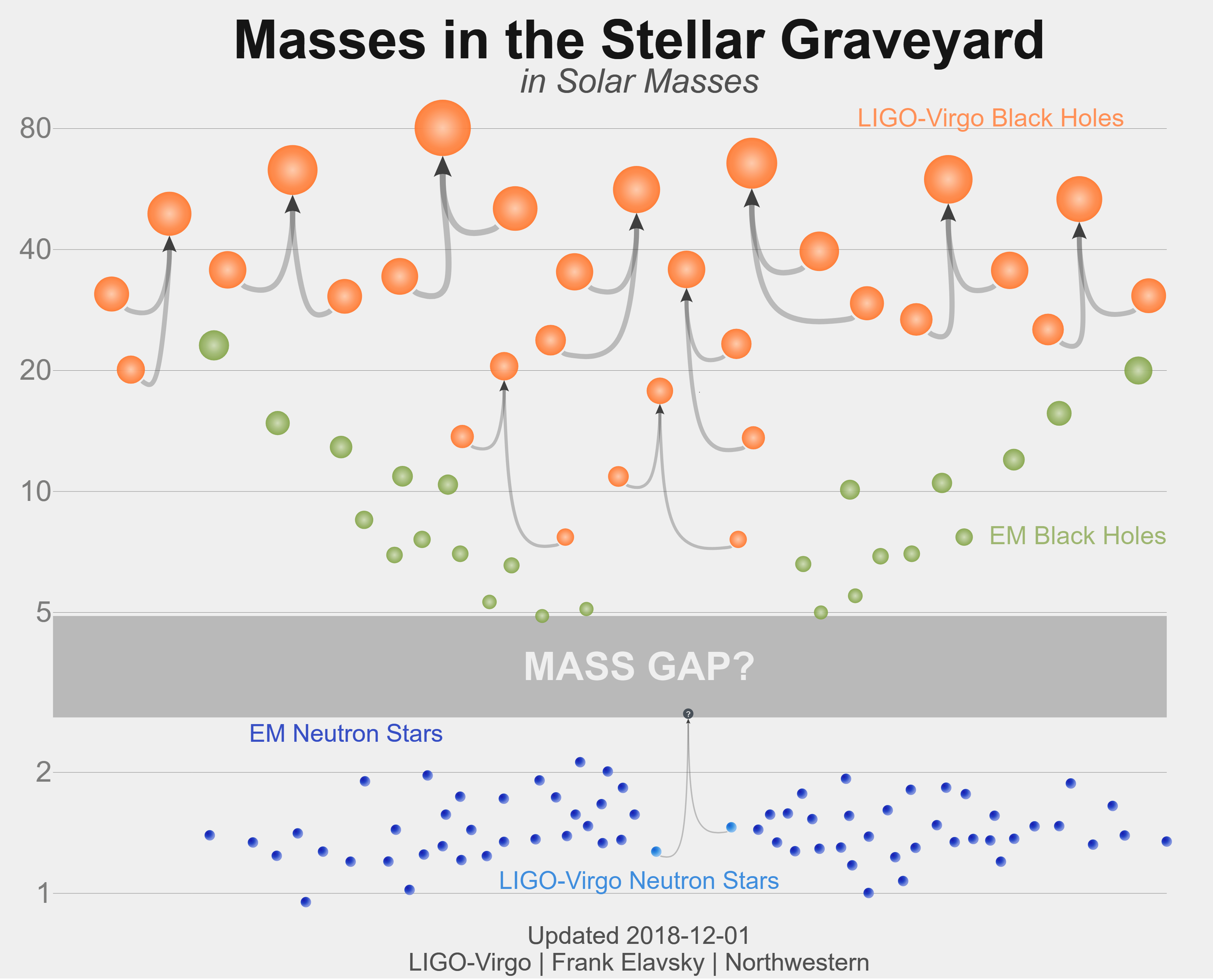}
\end{center}
\caption{\small\label{fig:HHmass} Distribution of stellar BH and NS masses, in units of $M_\odot$. The masses for BHs detected through electromagnetic observations are in green; the BHs measured by gravitational-wave observations are in orange.  Neutron stars measured with electromagnetic observations are in blue; the masses of the NSs that merged in GW170817 are in the center. 
Credit: LIGO-Virgo/Frank Elavsky/Northwestern}
\end{figure}

GW150914 is not the only binary BH merger observed by LIGO/Virgo. At the end of O2 runs, the Collaborations presented results on the mass, spin, and redshift distributions of the ten binary BH mergers detected in the first and second observing runs \cite{LVBH1}.
The key parameters of the observed events are reported in Table \ref{tab:summ}.
Until 2016, there had only been a couple dozens of stellar BHs indirectly detected via electromagnetic radiation, mainly X-rays. 
The largest of them was $\sim 20 M_\odot$; the more likely mass was 5-10 $ M_\odot$.
The common characteristic of almost all BHs reported using GWs is that the masses are larger than expected from previous observations and theoretical astrophysical models (biased by observations).
Fig. \ref{fig:HHmass} shows the distribution of the masses of stellar remnants measured in many different ways. 
Each observation through the merger of binary systems corresponds to three objects: the individual two BHs before merging, and the final state. Only GW170608, the lightest binary BH system, seems matching the pre-discovery prejudice about BH masses. 

The simple distribution of masses (initial and final) in the Table \ref{tab:summ} probably requires some revisions of astrophysical models of stellar evolution.
From these events, a merger rate of binary BH of (10-100) per Gpc$^{-3}$ y$^{-1}$  (at 90\% confidence level) is derived. 
The additional events recorded during the O3 run with a rate of about one event per week\footnote{The LIGO and Virgo Collaborations maintain an on-line service event database (GraceDB), which organizes candidate events from GW searches and provides an environment to record information about follow-ups. See \url{https://gracedb.ligo.org/}. } will soon reduce the uncertainty band of this figure. In addition, binary NS and BH+ NS merger rates will be provided.

The measured rate can be compared with theoretical and phenomenological models of binary BH formation, based on population synthesis \cite{LVBH2}. 
The population synthesis requires modelling of stellar evolution combined with the influence of their evolutionary environments. The large number of observed BHs with masses $> 10\ M_\odot$ is at present unexplained.
In addition, electromagnetic observations and modeling of systems containing BHs have led to speculation about the existence of potential gaps in the mass spectrum, between the most massive neutron stars ($\sim 2.1-2.5\ M_\odot$) and the lightest BHs ($\sim 5\ M_\odot$).
A wider gap seems to exist between stellar BHs and supermassive black holes (SMBHs, with mass of the order of hundreds of thousands, to billions of times, $ M_\odot$).

\section{Multimessenger perspectives after GW170817}
\label{sec:GWcosmo}

\subsection{Standard siren and observational cosmology}
GW170817 represents the first event for which both gravitational and electromagnetic waves from a single astrophysical source have been observed, thereby also opening new perspectives in fields other from astrophysics, as discussed in \cite{4}.
For instance, the combined observation was used to constrain the difference between the speed of light, $c$, and the speed of gravity, $v_{gw}$, by improving the previous estimate by about 14 orders of magnitude.
In addition, the observation of GW170817 allowed for investigation of the equivalence principle and Lorentz invariance.
\vskip 0.3cm
\noindent\textbf{Exercise:} {\small 
Using the information in Fig. \ref{fig:chirp2} and the measured luminosity distance $D_L=44$ Mpc of the source, show that the difference $v_{gw}-c$ is constrained to stay within $-3\times 10^{-15}c$ and $+7\times 10^{-16}c$.}
\vskip 0.3cm

In cosmology, GWs provide a novel approach to measuring the expansion of the Universe: the distance estimate using GWs is completely independent of the cosmic distance ladder derived from electromagnetic observations.
The coalescence of a binary NS system, in fact, represents a ``standard siren'', which is the gravitational analogue of the electromagnetic ``standard candle'': their intrinsic luminosity distance $D_L$ can be inferred directly from observations (masses of compact objects and other parameters of the system). 
If the source redshift is known, these information can be used to determine the Hubble constant $H_0$.
In the case of GW170817, the analysis of the waveform yielded $D_L= 44$ Mpc, assuming that the sky position of GW170817 was exactly coincident with its optical counterpart. The associated uncertainty on $H_0$ corresponds to $\sim 15\%$, resulting from a combination of instrumental noise in the detectors and the poor determination of the inclination of the orbital plane of the binary neutron star system with respect to the Earth. 
To estimate $H_0$, the luminosity distance to NGC4993 was combined with the galaxy's radial velocity, a quantity affected (after correction for the \textit{peculiar velocity} due to local irregularities and {}``clumpiness'') by the Hubble expansion. 

The obtained value of 70 km s$^{-1}$ Mpc$^{-1}$ \cite{hubble} can be compared with the two state-of-the-art analyses that solely use electromagnetic data: the analysis of cosmic microwave background radiation from the Planck satellite and the SHoES analysis combining the Cepheid variable and type Ia supernovae data from the relatively nearby universe. The Planck and SHoES results are not in agreement with each other at the level of more than 4$\sigma$ \cite{riess}. 
Due to the large amount of uncertainty on this particular GW measurement, the derived $H_0$ is consistent with both the Planck and SHoES values.
However, this marks an important milestone in the fundamental problem of measuring the expansion rate of the Universe, and future GW observations will be able to make increasingly precise measurements of this quantity.

\subsection{Jets and neutrinos}
\label{sec:neutrino}

X-ray and radio emission were discovered at the AT 2017gfo (the optical position corresponding to GW170817) about 9 and 16 days after the merger, respectively. 
Both the X-ray and radio emission likely arise from a physical process that is distinct from the one that generates the UV/optical/near-infrared emission discussed in \S \ref{sec:at17}.

The most plausible model for the delayed X-ray and radio afterglow emission, consistent with the kilonova description of the NS merger as proposed in \cite{Metzger}, is the presence of an off-axis jet, that is, pointing away from Earth.
The details are still not completely determined. The delayed X-ray and radio production are consistent with different scenarios: 
with the presence or a simple uniform jet observed at Earth;
or with the presence of a more complex, structured jet in which the energy decreases with the angular distance from the axis; 
or with the presence of a \textit{cocoon} accelerated  quasi-isotropically at mildly relativistic velocities by the jet.
In all cases, the Earth location is a relatively large angle $\theta_v$ with respect to the jet axis, with a value in agreement with the GW observation given by Eq. (\ref{eq:ns4}).

\begin{figure}[tb]
\begin{center}
\includegraphics[width=12.0cm]{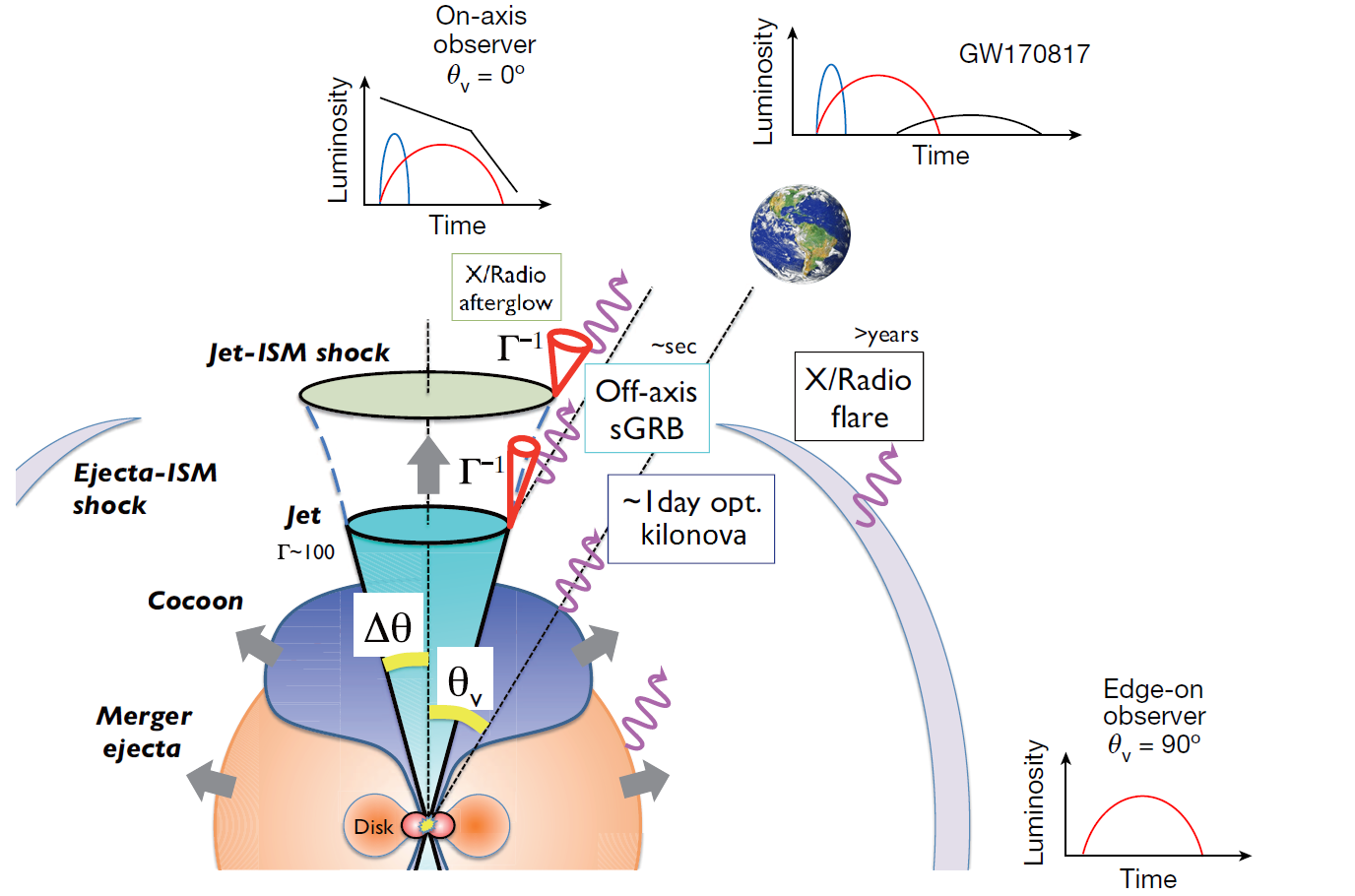}
\end{center}
\caption{\small\label{fig:sketch} 
Sketch of the geometry of GW170817 and production of electromagnetic transients. See text for details. Adapted from \cite{Metzger,Troja,Ioka}.
}
\end{figure}

Referring to Fig. \ref{fig:sketch}, the collimated jet (black solid cone) emits synchrotron radiation visible at radio, X-ray and optical wavelengths. This afterglow emission (black line in the luminosity vs. time plots on top of the figure) outshines all other components if the jet is seen on-axis. However, to an off-axis observer, the afterglow emission appears as a low-luminosity component delayed by several days or weeks (luminosity in top-right plot).
The jet opening angle, $\theta_v$, is related to the Lorentz $\Gamma$ factor of the particles in the jet as $\theta_v=1/\Gamma$. As the jet slows, the opening angle broaden. 
Following the NS merger, a fast-moving \textit{merger ejecta} with speed $\sim 0.2 c$ and neutron-rich (orange shells) emits an isotropic kilonova peaking in the infrared (red lines in the luminosity-time plots).
Edge-on observations ($\theta_v\sim 90^\circ$) detect only this component.
A larger mass neutron-free wind (\textit{cocoon}) along the polar axis (blue arrows) produces an emission (blue lines in the luminosity-time plots) peaking at optical wavelengths. This emission, although isotropic, is not visible to edge-on observers because it is only visible within a range of angles and otherwise shielded by the high-opacity ejecta.

In a GRB, neutrinos and $\gamma$-rays are expected to be produced by the central engine's activity, that results in fluctuations of the relativistic outflow, creating internal shocks in the ejecta. 
These internal shocks accelerate electrons and protons in the outflow through the process known as \textit{Fermi acceleration} (Chapter 6 of \cite{spurio}). 
Shock-accelerated electrons radiate their energy through synchrotron or inverse-Compton radiation, producing $\gamma$-rays. 
Shock-accelerated protons interact with ambient photons and $\gamma$-rays ($p\gamma$ process) as well as with other, non-relativistic protons ($pp$ process), producing charged pions and kaons. 
Secondary pions and kaons decay into high-energy neutrinos through: 
\begin{equation}\label{eq:nu1}
\pi^\pm ,\ k^\pm \rightarrow \mu^\pm + \nu_\mu (\overline \nu_\mu)  \end{equation}
Since internal shocks in the relativistic outflow result in both $\gamma$-ray and high-energy neutrinos, the latter are expected to be produced at the same time of the GRB emission. In addition, since efficient production of neutrinos requires high target density of radiation and/or particles, typical neutrino production is likely to take place close to the central engine.

The radiation observed in the afterglow phase is mainly produced by synchrotron emission of shock-accelerated electrons. The energy distribution of protons is expected to be similar to that of electrons.
Therefore, the softer emission spectrum during the afterglows indicate lower proton energies and lower neutrino production probability. 
However, because of the longer time for Fermi acceleration, some models foresee that GRBs can accelerate protons to energies up to $10^{11}$ GeV. This corresponds to the maximum energy of observed charged cosmic rays. 
Consequently, few ultra-high-energy neutrinos of energies $\sim 10^8-10^9$ GeV might be emitted during the afterglow phase. 

\begin{figure}[tb]
\begin{center}
\includegraphics[width=12.3cm]{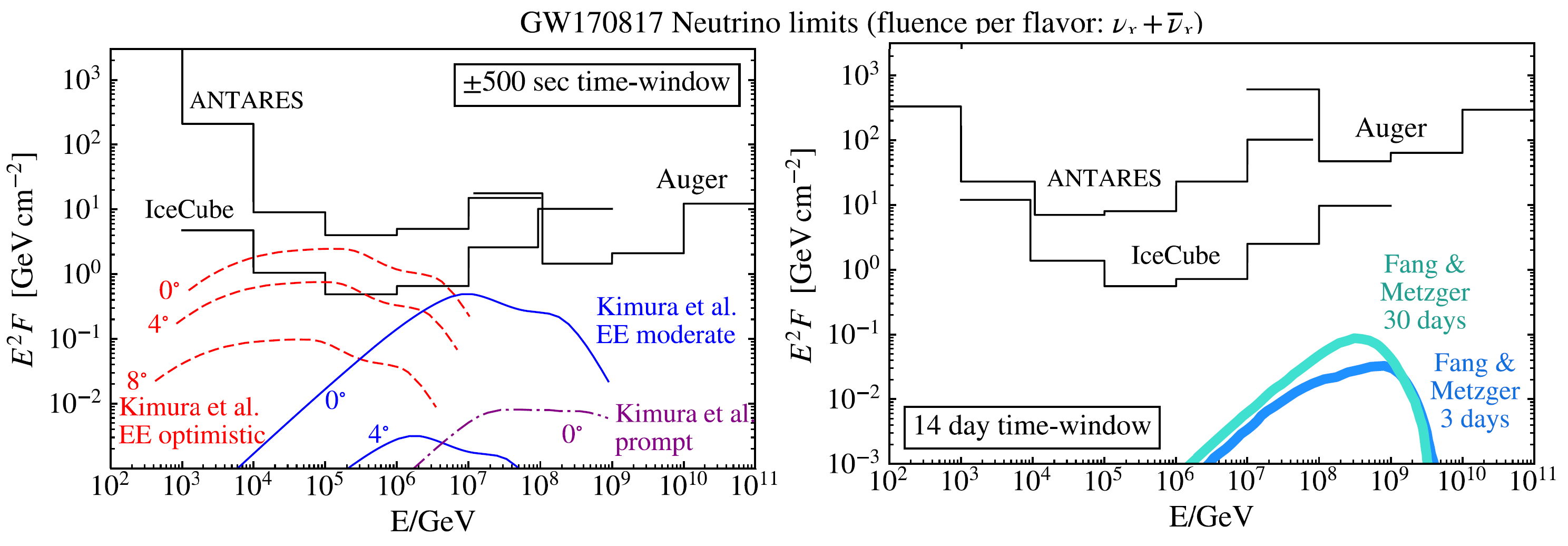}
\end{center}
\caption{\small\label{fig:GWnus} Upper limits (at 90\% confidence level) on the neutrino (sum of $\nu_x$ and $\overline\nu_x$ of all flavors) spectral fluence (i.e. energy flux, GeV cm$^{-2}$) from GW170817 during a $\pm$500 s window centered on the GW trigger time (left panel), and a 14 day window following the GW trigger (right panel). For each detector (ANTARES and IceCube neutrino telescopes; Auger extensive air shower array), limits are calculated separately for each energy decade, assuming a spectral fluence $F(E) = F_0\cdot [E/GeV]^{-2}$ in that decade only. Also shown are predictions by neutrino emission models: the models from \cite{Kimura} scaled to a distance of 40 Mpc and shown for the case of three different viewing angles. In the right plot, the models from \cite{FangMetzger}. From \cite{GW17nu}. }
\end{figure}

In the case of the off-axis scenario of GW170817, the active neutrino telescopes (ANTARES, IceCube, see: Chap. 10 of \cite{spurio}) and the Pierre Auger large air shower array (Chap. 7 of \cite{spurio}) searched for high-energy neutrino emission in a time windows of $\pm 500$ s from the coalescence time \cite{GW17nu}. 
The most promising neutrino-production mechanism seems to be related to the extended $\gamma$-ray emission phase during the afterglow: the (relatively) low Lorentz factor of the expanding material results in high meson production efficiency. 
The models for the neutrino flux associated with the prompt GRB emission seems to be less favourable for neutrino production.
Finally, a search extended to 14 days after the merger was also performed to account for neutrino produced at the end of a (possible) acceleration of protons up to the highest energies. In all cases, no neutrino candidates have been found, Fig. \ref{fig:GWnus}.

\subsection{Bursts of GWs from Stellar Gravitational Collapses\label{sec:GWbursts}}
Neutron stars and stellar black holes are formed from the core collapse of an accreting white dwarf or the gravitational collapse of a highly massive star. 
There is increasing evidence that some gravitational collapses (hypernovae and collapsars) also produce most of the observed long gamma-ray bursts. Details provided in Chap. 12 of \cite{spurio}.
Many pulsars present large measured speeds relative to their neighbors: this observation suggests that some supernovae do exhibit substantial non-spherical motion, perhaps because of dynamical instabilities in rapidly rotating, massive progenitor stars. 
If the collapse occurs non-spherically (a spherically symmetric explosion or implosion does not have a quadrupole moment), gravitational waves are produced. 

However, core collapse supernovae produce bursts whose time evolution is theoretically poorly known. Although computer simulations are available, predictions of strains $h$ of the produced gravitational waves remain subject to large uncertainties.
Thus, although algorithms for searching for bursts of GWs have been developed, they are necessarily less sensitive than matched-filter approaches, in which known phase evolution can be exploited. 

To make a rough estimate of the amplitude $h$ (following \cite{14Sa09}), we can start from the energy flux carried by the GW.
If the GW burst can be approximated with a triangular shape, with a linear increase of the strain $h$ from 0 to $h_\circ$ and a similar decrease to 0 in a time window $\Delta t\sim 1-10$ ms, then 
\begin{equation}\label{eq:gw1}
\dot{h} \simeq 
2h_\circ/\Delta t 
\simeq h_\circ \nu_{gw} \ , 
\end{equation}
where we used the fact that the characteristic frequency of a GW signal produced in a time interval is $ \nu_{gw} \simeq 2/\Delta t$.  
Finally, using the relation for the luminosity distance (\ref{eq:dl1}), assuming that ${\mathcal L} = E/\Delta t$, and from the energy flux Eq. (\ref{eq:enegw1}), we obtain (neglecting the numerical factor)
\begin{equation}\label{eq:gw2}
h_\circ \simeq \frac{(G/c^3)^{1/2}}{D_L \nu_{gw}} \sqrt{\frac{E}{\Delta T}} \ .
\end{equation}
Here, $E$ is the total energy radiated as gravitational waves. According to simulations, for a massive star of ten solar masses, $E\sim (10^{-7} - 10^{-5}) \times 10 M_\odot$. 

Using representative values for a supernova burst in the central region of our Galaxy, at $D_L=10$ kpc, lasting for 1 ms, emitting the (conservative) energy equivalent of $10^{-6} M_\odot$ at a frequency of 1 kHz, the strain amplitude (\ref{eq:gw2}) would be
\begin{equation}\label{eq:gw3}
h_\circ \simeq 10^{-20} \biggl( \frac{E}{10^{-6} M_\odot}\biggr)^{1/2}
\biggl( \frac{\textrm{1 ms}}{\Delta t}\biggr) ^{1/2}
\biggl( \frac{\textrm{1 kHz}}{\nu_{gw}}\biggr)
\biggl( \frac{\textrm{10 kpc}}{D_L}\biggr) \ .
\end{equation}
This amplitude is large enough for current ground-based detectors to observe a galactic supernova with a reasonably high confidence.
The event rate within 10 kpc is expected to be far too small to make an early detection likely.
Supernovae of Type II are believed to occur at a rate of 3 per century in a galaxy similar to the Milky Way. The Virgo supercluster has diameter of $\sim 30$ Mpc and contains about 2,500 large galaxies, thus we might expect an event rate of $\sim$ 50 per year. Hypernova events are considerably rarer.
Note that until one reaches the Andromeda galaxy ($\sim$ 800 kpc), there is relatively little additional stellar mass beyond the edge of the Milky Way: nearby dwarf galaxies (such as the Large and Small Magellanic Clouds) contribute only a few percent additional mass. At the distance of Andromeda, the strain (\ref{eq:gw3}) decreases by about two orders of magnitude with respect to a Galactic event.

Due to the large uncertainties about modeling stain amplitudes, an easy detection of gravitational waves from SN bursts over a short timescale seems difficult. Maybe the multimessenger astrophysics approach can work in the opposite direction in this case: if a supernova burst is observed optically or in neutrino observatories in a galaxy sufficiently close to us, the gravitational wave imprint can be extracted off-line from the data. 
This requires the localization in space of the event and, with a greater degree of difficulty, a temporal localization of the event. In this case, we can learn about the burst mechanism by analyzing the gravitational wave strain $h(t)$.

\section{Conclusions and outlook}
\label{sec:conclusion}

Gravitational waves are travelling ripples in space-time, generated when heavy cosmic objects accelerate. These distortions, described as waves, move outward from the source at the speed of light: after a century of debate and searches, on September 2015 the strain of a GW was recorded by the LIGO laser interferometer. 
This first direct observation of a GW is a milestone, not only for providing a means to investigate general relativity in a previously inaccessible regime. In fact, GWs allows exploring the distant non-thermal Universe in a way completely independent by the electromagnetic radiation. The great opportunities opened by GW detections have been made evident in August 2017 with the arrival of the GW induced by the coalescence of a binary NS system (GW170817).
As an example of the new insights in physics, astrophysics and cosmology covered by the new instruments for the detection of GWs are:

\noindent $\bullet$ direct observation that GW carries energy and the measurement of their propagation speed;

\noindent $\bullet$ tests of general relativity, also under extreme strong-field conditions;

\noindent $\bullet$ direct observation of black holes in binary systems, including the measurement of their masses and a test of the fundamental \textit{no-hair} theorem;

\noindent $\bullet$ information on the equation of state and other properties of neutron stars;

\noindent $\bullet$ measurement of the Hubble constant and the definition of a new ``astronomy distance ladder'' with a completely different technique; 

\noindent $\bullet$ confirmation of the origin of short gamma-ray burst by coalescence of neutron stars;

\noindent $\bullet$ study of the models for the accretion disks and jets;

\noindent $\bullet$ insights into the earliest stages of the evolution of the Universe through primordial GWs;

\noindent $\bullet$ studies of galactic merging through the observation of coalescing massive black holes at their centers.

The study of the Universe with probes different from the electromagnetic radiation has only recently reached its maturity. 
The joint effort to understand high-energy astrophysics phenomena using cosmic rays, $\gamma$-rays, neutrinos, gravitational waves, in addition to electromagnetic radiation, is the aim of \textit{multimessenger astrophysics}.
The goals of future astroparticle experiments include not only astrophysics, but also studies related to particle physics, general physics and cosmology. This includes, for instance: 
the search in the cosmic radiation for particles not included in the Standard Model (the \textit{dark matter} problem); 
the measurement of the neutrino mass hierarchy and of the possible CP violation in the leptonic sector; 
the measurement of particle's (protons, photons, neutrinos) cross-sections at energies unattainable in Earth-bound accelerators; 
the search for baryon number violation;
the measurement of the extragalactic background light using the
attenuation of $\gamma$-rays;
the understanding of the cosmic history of star formation; 
the search for hints on the origin of the matter-antimatter asymmetry of the Universe; 
the exploration of the fundamental nature of space-time and of the vacuum (the \textit{dark energy} problem), ...

\vskip 0.2cm
The first half part of the twentieth century saw a strict interconnection of particle and astroparticle physics. The advent of accelerators decoupled the two fields in the second part of the century. 
From the 1950s, the study of the microcosm had an impressive growth, forced by the increasing energy of accelerators from the MeV to the TeV scale. 
To go far beyond the energy scale (10 TeV) reached by the LHC, efforts probably at the limit of human (financial) possibilities are required.
The return to the use of cosmic accelerators will probably be a necessity. From the particle physics point of view, the possibility of using cosmic beams to improve our understanding of Nature will depend upon either the detailed understanding of cosmic acceleration and on the development of methods for controlling systematic errors introduced by our lack of understanding of these processes. 
Thus, the combined information arising from gravitational waves, from the measurements of $\gamma$-rays with high-resolution instruments, from high-statistics measurements of charged CRs and from neutrino telescopes is mandatory. 
For this purpose, multimessenger observations are not just an advantage, but a necessity.

\appendix

\section{Sensitivity of ground-based interferometers}
\label{sec:sensitivity}

\begin{figure}[tb]
\begin{center}
\includegraphics[width=11.0cm]{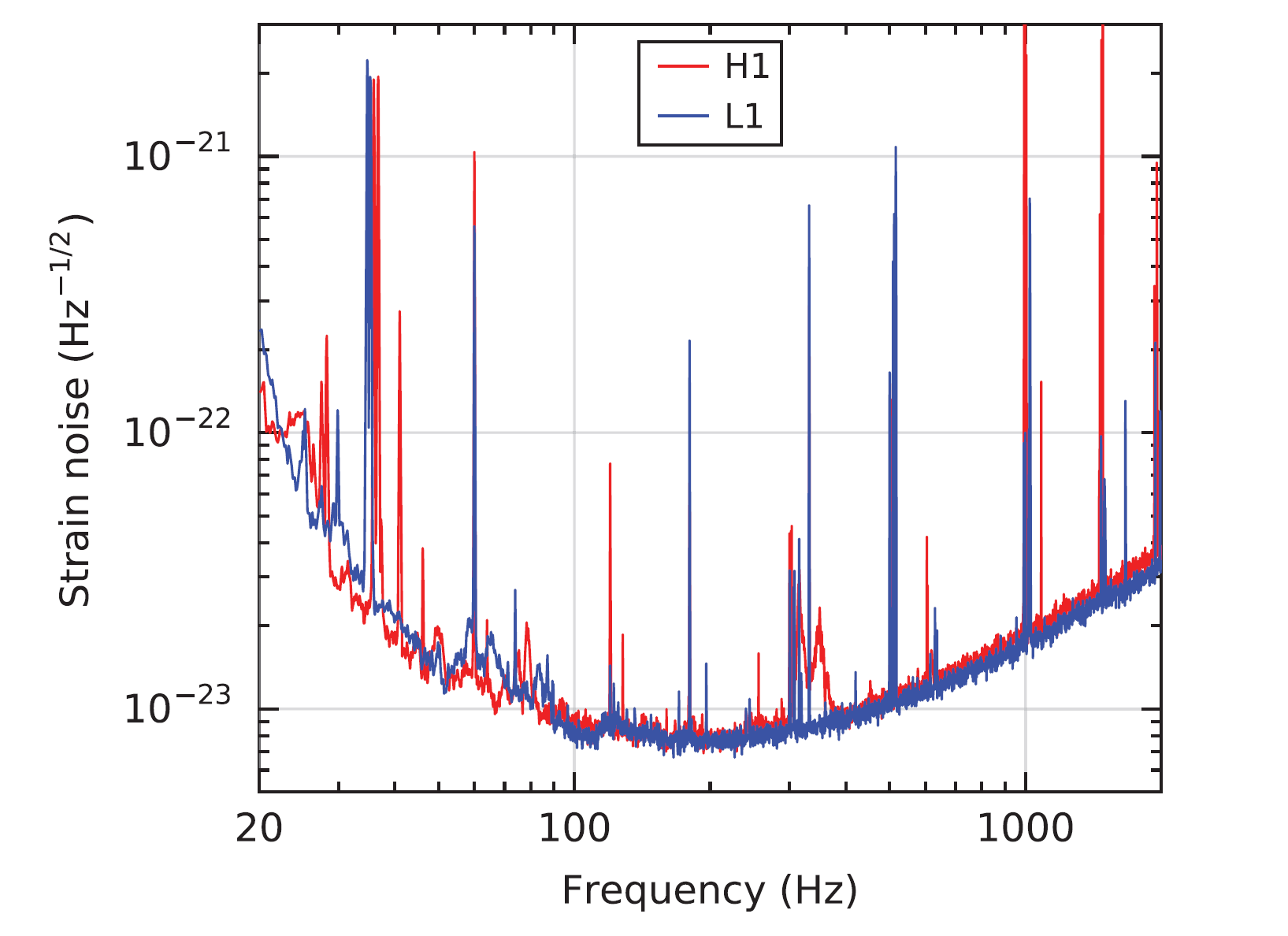}
\end{center}
\caption{\small\label{fig:noise} The aLIGO instrument noise at Hanford, WA (H1) and Livingston, LA (L1) at the time of GW150914. On the $y$-axis there is an amplitude spectral density, expressed in terms of equivalent gravitational-wave strain amplitude. The sensitivity is limited by photon shot noise at frequencies above 150 Hz, and by a superposition of other noise sources at lower frequencies. Narrow-band features include calibration lines (33-38, 330, and 1080 Hz), vibrational modes of suspension fibers (500 Hz and harmonics), and 60 Hz electric power grid harmonics. From \cite{1}.}
\end{figure}

A genuine GW signal must be extracted from the large background due to noise sources. These noise sources can be divided into two categories: \textbf{displacement noises}, as the thermal noise, the ground vibrations and the gravity gradient noises; and \textbf{sensing noises}, as the shot noise and the quantum effects, associated with the conversion of a small displacement into a readout signal.
The sensitivity of the detectors at different frequencies are represented with plots like Fig. \ref{fig:noise}.

The {thermal noise}, collective modes of motion of components of the apparatus, represents a generalization of Brownian motion, which arises from a coupling of a macroscopic element to its environment.  Interferometers perform measurements at frequencies far from the resonant frequencies (pendulum suspensions in few Hz range; internal vibrations of the mirrors at several kH) where the amplitude of thermal vibrations is largest. Thermal effects produce also other disturbances. Some of the mirrors (as the beam splitters) are partly transmissive and they absorb a small amount of light power during transmission. This absorption raises the temperature of the mirror and changes its index of refraction. The effect degrades the optical properties of the system, and effectively limits the amount of laser power that can be used in the detector. 

The {ground vibrations} are due to: mirrors, Earth's seismic background, man-made sources such as traffic and machinery, and wind and rain coupling to the ground through trees and buildings.
As interferometers bounce light forth and back between mirrors, each reflection introduces further vibrational noise. Suspension/isolation systems based on pendulums are used to reduce vibrations. 
A pendulum represents a mechanical filter for frequencies above its natural frequency. By hanging the mirrors on pendulums of 0.5-1.0 m length, filtering above a few Hertz are achieved. 

The {gravity gradient noise} is due to changes in the local Newtonian gravitational field, producing local tides on the timescale of the measurements, and cannot be screened out.  
For instance, seismic waves are accompanied by changes in the gravitational field, and changes in air pressure are accompanied by changes in air density. In addition, there are also environmental noises coming from man-made sources (traffic): overall, the spectrum of gravity gradient noise falls steeply with increasing frequency.

Among sensing noises, the {shot noise} occurs because the photons used for interferometry are quantized: light arriving at the beamsplitter in bunches on $N$ photons will be subject to Poisson statistics with uncertainty decreasing (for large $N$) as $\sqrt N$. Thus, shot noise is minimized by maximizing the photon arrival rate, or equivalently, the laser power. On the other hand, as the laser power is increased, the position sensing accuracy improves, with a final limit due to the Heisenberg uncertainty principle: the momentum transferred to the mirror by the measurement leads to a disturbance that can mask a gravitational wave (the \textbf{quantum effects} noise).

It is instructive from a didactic point of view to estimate the sensitivity limits on the strain $h$ due to the aforementioned shot noise. 
Let consider a laser bunch of $N$ photons of wavelength $\lambda$ and wave number $k=2\pi/\lambda$.
The uncertainty on the distance $\Delta x$ arising from a wave packet formalism is 
$$\Delta x \cdot k= \frac{1}{\sqrt N}$$ 
or, equivalently,
\begin{equation}\label{eq:wp1}
\Delta x \simeq \frac{\lambda}{2\pi\sqrt N}
\end{equation}
To measure a GW with frequency $\nu_{gw}$, one has to make at least $2\nu_{gw}$ measurements per second, so one can accumulate photons for a time $\Delta t_{gw}$ such as:
\begin{equation}\label{eq:wp2}
\Delta t_{gw} \simeq \frac{1}{2\nu_{gw}} \ .
\end{equation}
If we use a laser with power $P$ (Watt) with photons of energy 
$E_\gamma =hc/\lambda=2\pi \hbar c/\lambda$ (we will use the reduced Plank constant $\hbar$ to avoid confusion with the strain $h$), the number of photons $N$ in a bunch of length $\Delta t_{gw}$ is
\begin{equation}\label{eq:wp3}
N = \frac{P\cdot \Delta t_{gw}}{E_\gamma }= 
\frac{ P\cdot \Delta t_{gw}\cdot \lambda}{2\pi\hbar c} = 
\frac{ P\cdot  \lambda}{4\pi\cdot \hbar  c \cdot \nu_{gw}}
\end{equation}
The strain $h$ from a GW induces a variation $\Delta L$ on the test masses that depends on the interferometer length $L$ and on the number of reflections of the laser light $n_{rif}$ in the Fabry-Perot cavities (see \S \ref{sec:experiments}):
\begin{equation}\label{eq:wp4}
\Delta L = h \cdot n_{rif}\cdot L \quad \longrightarrow \quad
h= \frac{\Delta L} {n_{rif}L} \ ,
\end{equation}
equivalent to Eq. (\ref{eq:bygw11b}) when the reflections are taken into account. 
The sensibility of the interferometer to the strain $h$ above the shot noise corresponds to the condition $\Delta L \gtrsim \Delta x$, where $\Delta x$ is given by Eq. (\ref{eq:wp1}).  
Thus, the in order to overcome this noise, the value of $h$ in the interferometer must be:
\begin{equation}\label{eq:wp5}
h\gtrsim \frac{\lambda}{2\pi\cdot n_{rif}\cdot L\cdot \sqrt N}   
\end{equation}
or, using (\ref{eq:wp3}),
\begin{equation}\label{eq:wp6}
h\gtrsim \frac{1}{n_{rif}\cdot L}\sqrt{ \frac{\hbar c \cdot \lambda }{\pi} \cdot \frac{\nu_{gw}}{P}  } \ .
\end{equation}
To insert numerical values into (\ref{eq:wp6}), we refer to parameters in the O1 run and reported in Fig. \ref{fig:LIGO}. First, we rearrange the expression assuming $n_{rif}=270$, $P=22$ W and a GW frequency $\nu_{gw}=100$ Hz:
\begin{equation}\label{eq:wp7}
h\gtrsim \frac{100^{1/2} }{270 \cdot 22^{1/2}} \sqrt{ \frac{ \hbar c \cdot \lambda }{\pi L^2} }
\cdot \biggl( \frac{270}{n_{rif}} \biggr)
\cdot \biggl( \frac{25 \textrm{ W}}{P} \biggr)^{1/2}
\cdot \biggl( \frac{\nu_{gw}}{100 \textrm{ Hz}} \biggr)^{1/2}
\end{equation}
and, after inserting numerical constants ($c=3\times 10^8$ m/s, $\hbar =1.05\times 10^{34}$ J$\cdot$s) and parameter values for LIGO 
($\lambda =1064$ nm, $L=4\times 10^3$ m), we obtain
\begin{equation}\label{eq:wp8}
h\gtrsim  {4\times 10^{-22}}  
\cdot \biggl( \frac{22 \textrm{ W}}{P} \biggr)^{1/2}
\cdot \biggl( \frac{\nu_{gw}}{100 \textrm{ Hz}} \biggr)^{1/2}
\end{equation}
The recycling of the laser light in aLIGO increases the power of the laser with a gain factor of $\sim 40$, thus reducing the sensibility to a further ${40}^{-1/2}$, i.e.
\begin{equation}\label{eq:wp9}
h\gtrsim  \frac{4\times 10^{-22}} {\sqrt{40}}=6\times 10^{-23} 
\cdot \biggl( \frac{\nu_{gw}}{100 \textrm{ Hz}} \biggr)^{1/2} \ .
\end{equation}
Compare with the value on Fig. \ref{fig:noise} at the frequency of 100 Hz.

\begin{figure}[t]
\begin{center}
\includegraphics[width=12.3cm]{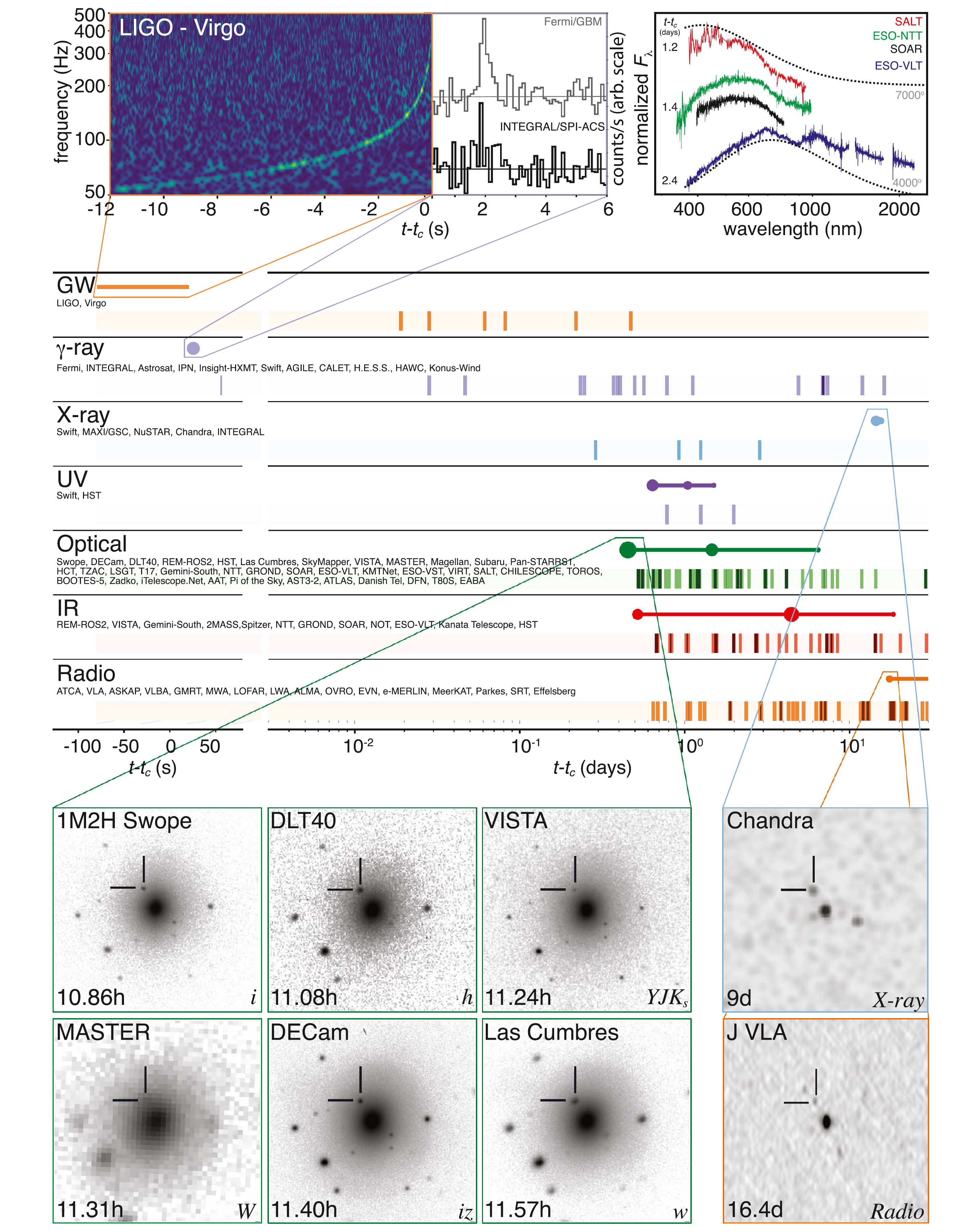}
\end{center}
\caption{\label{fig:1gw}
One of the symbols of multimessenger astrophysics \cite{3}: the timeline of the discovery of GW170817, GRB 170817A, and AT 2017gfo. All observations are shown by messenger and wavelength relative to the time $t_c$ of the gravitational-wave event. See text for details. }
\end{figure}


\end{document}